\documentclass[preprint,superscriptaddress,floatfix,sort&compress]{revtex4}

\usepackage[dvips]{graphicx}
\usepackage{amssymb,amsfonts,amsmath}
\usepackage{color}
\usepackage{capt-of}
\usepackage[font=scriptsize]{caption}

\usepackage{ulem}

\begin{document}

\title{Binding of bivalent transcription factors to active and inactive regions folds human chromosomes into loops, rosettes and domains}

\author{C.~A. Brackley}
\affiliation{SUPA, School of Physics \& Astronomy, University of Edinburgh, Mayfield Road, Edinburgh, EH9 3JZ, UK}
\author{J. Johnson}
\affiliation{SUPA, School of Physics \& Astronomy, University of Edinburgh, Mayfield Road, Edinburgh, EH9 3JZ, UK}
\author{S. Kelly}
\affiliation{Department of Plant Sciences, University of Oxford, South Parks Road, Oxford OX1 3RB, UK}
\author{P.~R. Cook}
\affiliation{Sir William Dunn School of Pathology, University of Oxford, South Parks Road, Oxford, OX1 3RE, UK}
\author{D. Marenduzzo}
\affiliation{SUPA, School of Physics \& Astronomy, University of Edinburgh, Mayfield Road, Edinburgh, EH9 3JZ, UK}

\begin{abstract}
{Biophysicists are modeling conformations of interphase chromosomes, often basing the strengths of interactions between segments distant on the genetic map on contact frequencies determined experimentally. Here, instead, we develop a fitting-free, minimal model: bivalent red and green ``transcription factors'' bind to cognate sites in runs of beads (``chromatin'') to form molecular bridges stabilizing loops. In the absence of additional explicit forces, molecular dynamic simulations reveal that bound ``factors'' spontaneously cluster -- red with red, green with green, but rarely red with green -- to give structures reminiscent of transcription factories. Binding of just two transcription factors (or proteins) to active and inactive regions of human chromosomes yields rosettes, topological domains, and contact maps much like those seen experimentally. This emergent ``bridging-induced attraction'' proves to be a robust, simple, and generic force able to organize interphase chromosomes at all scales.}
\end{abstract}

\maketitle

The conformations adopted by human chromosomes in 3D nuclear space are currently an important focus in genome biology, as they underlie gene activity in health, aging, and disease~\cite{Cavalli2013}. Chromosome conformation capture (3C) and high-throughput derivatives like ``Hi-C'' allow contacts between different chromatin segments to be mapped~\cite{Lieberman-Aiden2009}. Inspection of the resulting contact maps reveals some general principles, including: (i) Each chromosome folds into distinct ``topological domains'' during interphase (but not mitosis when transcription ceases); domains contain 0.1-2 Mbp, active and inactive regions tend to form separate domains, and sequences within a domain contact each other more often than those in different domains~\cite{Lieberman-Aiden2009,Dixon2012,Nora2012,Sexton2012,Naumova2013,Rao2014,Sexton2015}. (ii) Domains seem to be specified locally, as the same 20-Mbp region in a chromosomal fragment or the intact chromosome make much the same contacts~\cite{Rao2014}. (iii) Bound transcription factors like CTCF (the CCCTC-binding factor) and active transcription units are enriched at domain ``boundaries''~\cite{Dixon2012,Rao2014}. (iv) Factors bound to promoters and enhances stabilize loops~\cite{Simonis2006,Li2012,Jin2013,Zhang2013,Heidari2014,Rao2014,Mifsud2015}. (v) Co-regulated genes utilizing the same factors often contact each other when transcribed~\cite{Fullwood2009,Schoenfelder2010,Yaffe2011,Li2012,Papantonis2012}. (vi) Single-cell analyses show no two cells in the same population share exactly the same contacts, but the organization is non-random as certain contacts are seen more often than others~\cite{Nagano2013}. (vii) This organization is conserved; in budding yeast~\cite{Hsieh2015} and Caulobacter crescentus~\cite{Le2013}, ``chromosomal interaction domains'' (CIDs) are separated by strong promoters, and the bacterial ones are eliminated by inhibiting transcription. These principles point to central roles for transcription orchestrating this organization, with transcription factors providing the required specificity. 

Biophysicists are attempting to model this organization~\cite{Marenduzzo2006,Rosa2008,Nicodemi2008,Lieberman-Aiden2009,Duan2010,Junier2010,deVries2011,Kalhor2011,Rousseau2011,Barbieri2012,Bau2012,Brackley2013,Le2013,Naumova2013,Nagano2013,Benedetti2014,Jost2014,Tark-Dame2014,Trieu2014,Cheng2015,Hofmann2015,Johnson2015,Junier2015,Trussart2015,Zhang2015}, often basing the strength of interactions between segments distant in 1D sequence space on contact frequencies determined using Hi-C~\cite{Lieberman-Aiden2009,Duan2010,Kalhor2011,Rousseau2011,Nagano2013,Trieu2014,Giorgetti2014,Junier2015,Trussart2015,Zhang2015,Dekker2013,Serra2015}. To understand the principles underlying the organization, we use a minimal model without such fitting that was originally developed to analyze non-specific binding of proteins like histones to DNA~\cite{Brackley2013,Johnson2015}; here, we adapt it to include specific binding. Thus, spheres (representing transcription factors) bind briefly to cognate sites in runs of beads (representing chromatin) before dissociating. These factors provide an obvious connection with transcription, as they often associate with RNA polymerase (which can remain tightly bound to the template for $\sim$10 min as it transcribes the average human gene -- a binding that is also specific in the sense it occurs throughout a transcription unit but not elsewhere). [However, here, we only model transient binding.] Like many transcription factors (or complexes made up of several of these factors), ours are ``bivalent''; they can bind simultaneously to two or more segments of one fiber, to create molecular ``bridges'' that stabilize loops. More generally, our spheres could represent any bivalent DNA-binding complex that binds specifically. 

In contrast to previous work, our model is fitting free. Instead of beginning with experimentally-determined Hi-C data, we start with 1D information (i.e., whether a particular genomic region is transcriptionally active or not) and use it to generate a population of possible chromosome structures (considering fibers with more subunits than those used previously); only then, do we compare the resulting contacts with those seen experimentally. Remarkably, our coarse-grained molecular dynamic (MD) simulations show fibers spontaneously fold into structures possessing the key features outlined above. We uncover an emergent force that can act through the binding of just two (or more) types of transcription factor to their cognate sites that is able to organize interphase chromosomes locally and globally -- all without inclusion of any explicit attractive force between distant segments, or between factors. 

\section*{Results}

\subsection*{Chromatin fibers spontaneously assemble into imperfect rosettes}

Our MD simulations use the LAMMPS software package (Large-scale Atomic/Molecular Massively Parallel Simulator~\cite{Plimpton1995}) run in Brownian dynamics (BD) mode (see Supporting Information for more details). We begin with a ``chromatin fiber'' of 5,000 30-nm beads -- representing 15-Mbp -- diffusing amongst 30-nm ``transcription factors'' ({Fig. 1A}). Initially, ``transcription factors'' (hereafter factors) have no affinity for any bead in the fiber (which follows a self-avoiding random walk), but then binding is ``switched'' on so they now have a high affinity for every 20$^{\rm th}$ bead (pink), and a low affinity for all others (emulating the tight binding of transcription factors to cognate sites and non-specific binding elsewhere). Importantly, factors can bind to two (or more) beads, and affinities are just large enough to favor binding. Consequently, a factor often binds to a low-affinity site, dissociates, and rebinds nearby. As this process repeats, the factor may reach a high-affinity site and remain bound long enough to stabilize a loop ({Fig. 1A}); bound factors now spontaneously cluster ({Fig. 1B}, {C}; {Movies S1} and {S2}). The force driving analogous clustering after non-specific binding was dubbed the ``bridging-induced attraction''; it operates even though no explicit attraction between factors or between beads was specified, and it was not seen with monovalent factors or irreversible binding~\cite{Brackley2013}. Earlier work also shows such clustering occurs with 20-nm beads~\cite{Brackley2013}, so the results we now present should also apply to chromatin fibers of this (or different) size.

\begin{figure}
\centerline{\includegraphics[width=11.cm]{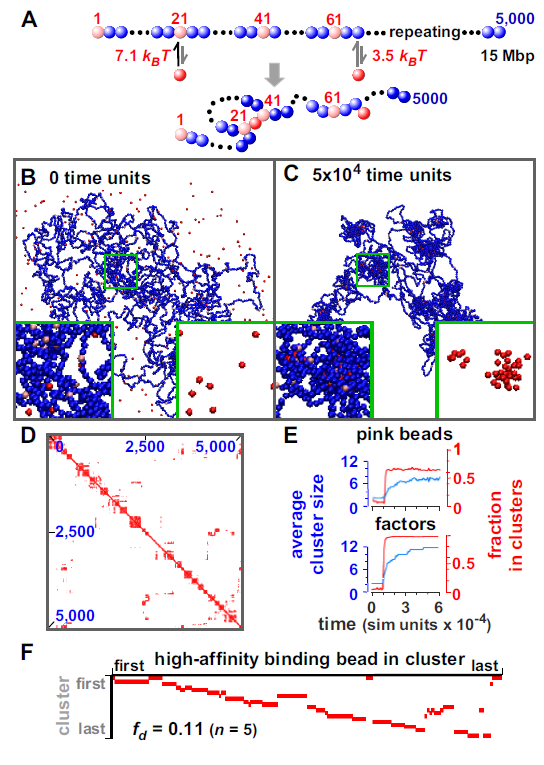}}
\caption{
{\noindent{\bf Bound ``factors'' spontaneously cluster.}
(A) Overview. MD simulations involved a 3-$\mu$m cube containing 
250 30-nm red spheres (``transcription factors''; volume fraction 0.01\% or ~15 nM), and 
a fiber of 5,000 30-nm beads (15-Mbp “chromatin”, so each bead contains 3 kbp; persistence 
length 90 nm; volume fraction 0.26\%, so chromatin is “dilute”). 
Every 20th bead is pink, others blue. Beads begin to interact (strength indicated) with factors 
after 10$^4$ time units if centers lie within 54 nm; here, binding of a factor to beads 21 and 41 
creates a loop.
(B,C) Snapshots after different times; insets show magnifications of boxed areas (with/without 
chromatin).
(D) Contact map after 5$\times$10$^4$ time units (axes give bead numbers; data from one run). Here (and
unless stated otherwise), a contact is scored if bead centers lie within 150 nm, and contacts made
by 40 adjacent beads are binned; a red pixel then marks contacts between beads at positions
indicated, with intensity (white to red) reflecting contact number (low to high). Blocks along the
diagonal mark many contacts made by clusters of bound factors.
(E) Average cluster size, and fraction in clusters, for pink beads and factors (data sampled every
1,000 time units). Two or more pink beads are in one cluster if centers lie $<$90 nm apart. Small
clusters form quickly, and slowly enlarge to the steady-state size.
(F) Rosettogram. A red pixel marks the presence of a high-affinity bead in a cluster; increasing
numbers of abutting pixels in one row reflect increasing numbers of loops in a rosette involving
near-neighbor high-affinity sites. Most clusters contain $\ge$2 loops. fd: disorganized fraction
(average of 5 runs).}}
\end{figure}

We next examine some properties of the system. As binding compacts the fiber, and as beads in/around each cluster make many contacts, blocks of red pixels are seen along the diagonal in the resulting contact map ({Fig. 1D}) -- as in Hi-C data~\cite{Lieberman-Aiden2009}. After switching on binding, clusters form in $<$1 min (one simulation time unit is 0.6 ms, calculated assuming a nuclear viscosity of 10 cP), and the fraction of pink beads in clusters increases rapidly ({Fig. 1E}). [We define two beads to be in the same cluster if centers lie within 90 nm.] Small clusters then slowly grow to reach a steady-state size with ~12 factors/cluster ({Fig. 1E}), when the entropic cost of gathering loops together (which scales nonlinearly with loop number~\cite{Duplantier1989}) limits further growth. It is likely that such ``coarsening'' is also dynamically hindered, as merging two clusters of loops (even when thermodynamically favored), would require passage over a free-energy barrier due to inter-loop interactions. Similar growth rates are found with all fibers described, and are not discussed further. 

As ``rosettes'' of loops are often found in models of chromosomes~\cite{Pienta1984,Cook1995}, we developed a suitable plot -- a ``rosettogram'' -- to assess how many existed in our simulations ({Fig. S1P}). In a rosettogram, there is a row for every cluster, and a column for every high-affinity bead in a cluster (other beads are not shown); then, a red pixel marks the presence of a binding bead in a cluster, and increasing numbers of abutting red pixels in a row reflect increasing numbers of loops (``petals'') involving near-neighbour high-affinity sites. In {Figure 1F}, the first cluster includes beads from both ends and two internal segments. However, the second organizes a ``perfect'' rosette with 14 petals around high-affinity beads 21, 41, 61, $\ldots$, 281; here, contacts display ``transitivity''~\cite{Rao2014}, with one loop running from bead 21 to 41, another from 41 to 61, and a third from 21 to 61 via 41. In contrast, ``overlapping loops'' (of the type running directly from bead 21 to 61, and from 41 to 81) are rarely seen here or in Hi-C data~\cite{Rao2014}. As most rows contain adjacent pixels, and as 80\% pink beads share a cluster with a nearest-neighbour pink bead, a measure of the disorganization (i.e., the disorganized fraction, $f_d$) is low ({Fig. S1}; {Table S1}). In other words, rosettes and local contacts are common. 

Clusters form if low-affinity sites are omitted ({Fig. S2A}); therefore, ``sliding'' from low- to high-affinity sites is not required for clustering. However, the contact map, rosettogram, and $f_d$ all point to a more disorganized structure. Randomly scattering the same number of high-affinity sites along a fiber creates more regular strings of rosettes ({Fig. S2B}), presumably because gaps between successive binding sites are exponentially distributed so that binding sites are naturally clustered nearer together in 1D genomic space (this is the so-called ``Poisson clumping''). Clusters and rosettes also form with a higher concentration of chromatin (i.e., in the ``semi-dilute'' regime~\cite{Doi1986}, see also below and {Fig. S5}).

\subsection*{Different transcription factors form into different clusters}

Binding of different factors to different beads was now analyzed. In {Figure 2Ai}, green factors interact only with light-green beads, and red ones only with pink beads. Again, no attraction is specified between factors, or between beads. Remarkably, clusters now contain only red factors -- or only green ones -- but rarely both (mixed clusters are not seen at the end of this simulation; {Fig. 2Aii}, { Movies S3}, { S4}). As before, clusters reach a steady state, but now with only $\sim$8.1 bound factors/cluster; the regular alternation of green and red binding sites renders cluster merging entropically more costly. Moreover, the contact map, rosettogram, and $f_d$ all point to a more disorganized structure than those seen previously; for example, there are now many ``overlapping'' loops where the fiber passes back and forth between a cluster of red factors to another with green ones ({Fig. 2Aiii},{ iv}). As expected, mixed clusters result if factors share binding sites ({Fig. S3}).

\begin{figure}
\centerline{\includegraphics[width=10.cm]{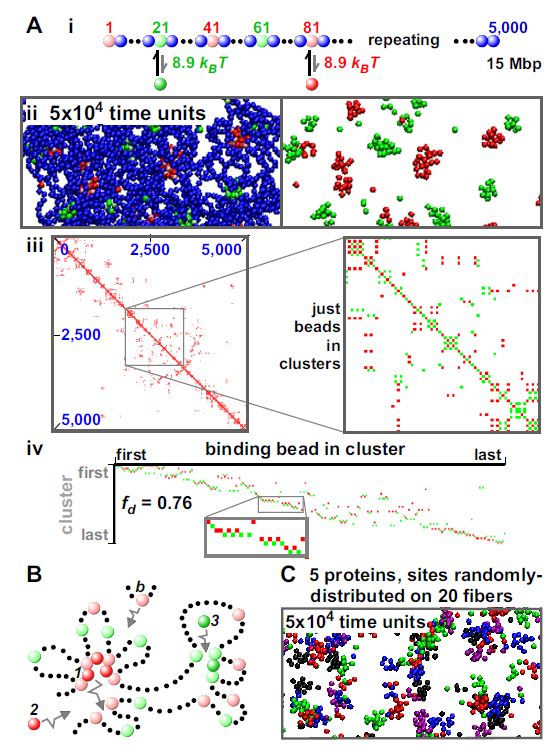}}
\caption{
{\bf Self-assembly into ``specialized'' clusters.} 
MD simulations were run as in Fig. 1, except for differences indicated.
(A) Red ($n=$250) and green ($n=$250) factors interact with pink and light-green beads,
respectively. (i) Binding beads at every 20th position alternate as indicated. (ii) Final snapshot of
central region (with/without chromatin); clusters contain either red or green factors. (iii) Final contact map;
blocks along the diagonal are small. The inset shows a high-resolution map involving only
binding beads in clusters; contacts are scored (without binning) if bead centers lie 90 nm apart
(not 150 nm), and any binding beads are treated as if they possess the color of factor binding
them. Here, red, green, and yellow pixels mark contacts between two pink beads, between two
light-green beads, and between a light-green and pink bead, respectively. Similarly-colored
pixels rarely abut in a row, as the fiber passes back and forth between differently-colored
clusters. (iv) Final rosettogram (pixels correspond to binding beads, and are colored as in the
contact map zoom); rows rarely contain abutting pixels of one color (reflected by a high fd).
(B) How ``specialized'' clusters form. See text.
(C) Red, green, dark-blue, purple, and black factors (500 of each) bind (7.1 kBT) to five sets of
cognate sites scattered randomly along 20 identical fibers (each with 2,000 beads representing 6
Mbp). The snapshot (taken after 5$\times$10$^4$ units; DNA not shown for clarity) shows that each factor
tends to cluster with similarly-colored ones. See also Figure S5.}
\end{figure}
 
Such self-organization into structures rich in certain DNA-binding proteins but not others is commonplace in nuclear biology (see Discussion). But what might drive this extraordinary self-assembly into ``specialized'' clusters in the absence of any explicit interaction between factors, or between beads? We suggest there are both entropic and kinetic drivers~\cite{Brackley2013}, and that the following one is important. Thus, early during the simulation in {Figure 2A}, a structure like that in {Figure 2B} might arise. Red protein $1$ is tightly bound to two pink beads, and so will rarely dissociate from the cluster; however, if it does it is likely to bind to a nearby pink bead (as there are so many). Further, as red protein $2$ and binding bead $b$ diffuse by, both are likely to join the same cluster (again because of the high local concentration of appropriate binding sites and factors). We are now in a positive feedback loop: the local concentration of red factors and pink beads makes it unlikely either will escape, and likely that more of both will be caught as they diffuse by. For the same reason, green protein $3$ is likely to bind to the right-hand cluster, and this cluster will tend to grow as other green factors and light-green sites are caught. Red and green clusters will inevitably be separate in 3D space because their cognate binding sites are separate in 1D sequence space, and cluster growth is limited when the entropic costs of bringing together ever-more loops becomes prohibitive.
	
In {Figure 2A}, red and green binding beads alternate, and overlapping loops pass back and forth between red and green clusters. Rosettes with many transitive loops result if 200-bead blocks containing 10 light-green beads (spaced every 20 beads) alternate with similar blocks containing pink beads ({Fig. S4}). As there are fewer binding beads of one color per block than the $\sim$12 often found in a cluster in the analogous simulation in {Figure 1}, successive blocks can form successive red and green clusters. Unsurprisingly, the 1D organization determines rosette and loop type. 

Distinct clusters also form if more factors and more fibers are introduced. In {Figures 2C} and { S5}, 5 differently-colored factors bind to distinct cognate sites scattered randomly along 20 fibers. Distinct clusters again form; 53\% contain factors of only one color, and in $>$80\% more than $>$80\% binding beads have the same color ({Fig. S5v}). Such clustering could underlie the high number of contacts seen between co-regulated genes that utilize the same factors~\cite{Fullwood2009,Schoenfelder2010,Yaffe2011,Li2012,Papantonis2012}. Inter-fiber contacts are rare ({Fig. S5iv}), as in Hi-C data~\cite{Lieberman-Aiden2009}. However, they constitute a higher fraction if just contacts made by binding beads in clusters are considered ({Fig. S5iv}). This is reminiscent of contacts made by active genomic regions; for example, most contacts made by (active) {\it SAMD4A} are inter-chromosomal (assessed by 4C~\cite{Papantonis2012}), as are most (active) sites binding estrogen receptor $\alpha$~\cite{Fullwood2009}. [{Figure S5} gives effects of the threshold used to define contacts on contact frequencies.]

The affinities of transcription factors for cognate sites are often tightly regulated, often by post-translational modification. Changing factor affinity was simulated using a fiber in which every 20$^{\rm th}$ bead was yellow ({Fig. S6}). Initially, red factors bind to yellow beads, and red clusters form. Then, we switch on an attraction between green factors and yellow beads that is stronger than the red-yellow attraction; consequently, green factors compete effectively with red ones, and red/green and all-green clusters develop. This provides a precedent for how one nuclear body (e.g., a transcription factory) might evolve into another.

\subsection*{Forming domains}

Topological domains are recognized as ``pyramids'' in contact maps prepared using data from Hi-C~\cite{Dixon2012,Nora2012,Sexton2015}) or simulations~\cite{Barbieri2012,Benedetti2014,Jost2014,Trussart2015,Zhang2015}. Our simulations demonstrate that the pattern of binding beads in our fibers determines whether pyramids are seen. For example, {Figure 3A} illustrates a partial contact map obtained using data from {Figure 1}. As clusters appear stochastically and tend to persist, a specified bead often clusters with different partners in different simulations. Then, pyramidal patterns, which are visible in a single run (analogous to Hi-C data from a single cell), become blurred on averaging results from progressively more simulations ({Fig. 3Aii}). [{Figure S7} gives complete contact maps for all simulations in {Figure 3}.]

\begin{figure}
\centerline{\includegraphics[width=10.cm]{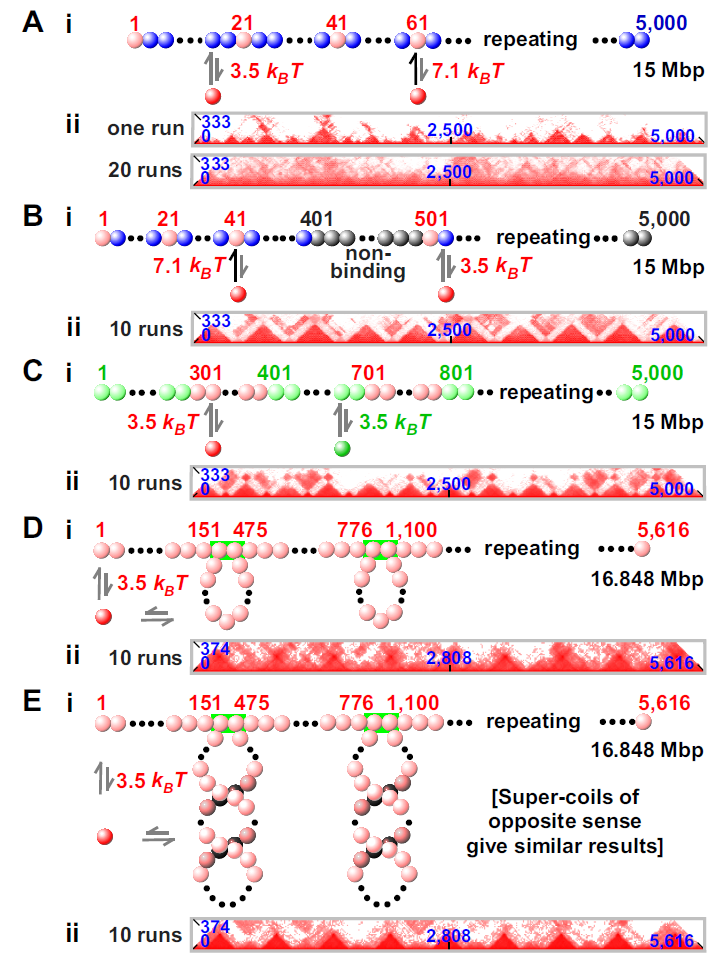}}
\caption{
{\bf Domain formation.} 
MD simulations were run as in Figure 1, unless stated otherwise ($n=$ total number of runs). In contact maps, only regions around the (horizontally-placed) diagonal are shown; axes give bead numbers (blue). [Figure S7 shows complete contact maps.]
(A) Clustering of factors does not necessarily lead to domains. (i) Red factors bind with high-affinity to every 20th bead (pink), and with low affinity to others (blue). (ii) Although pyramids are seen in the contact map after 1 run, averaging data from 20 runs blurs patterns.
(B) Gene deserts. (i) Blocks of 400 binding beads (blue and pink; every 20th bead is pink) alternate with blocks of 100 non-binding beads (grey); red factors bind to blue and pink beads with low and high affinity, respectively. (ii) Each pyramid coincides with a block of pink and blue beads, and is separated from the next by a disordered region.
(C) Hetero- and eu-chromatin. (i) Blocks of 300 light-green and 100 pink beads alternate; red and green factors bind to pink and light-green beads, respectively. (ii) Large pyramids alternate with small ones, reflecting reproducible assembly of blocks into domains.
(D) Loops. (i) The fiber is pre-organized into loops by forcing selected beads (green rectangles) to bind irreversibly; this results in 324-bead loops separated by 300 unlooped beads (plus 150 unlooped ones at each end). All beads are pink, and red factors can bind to any bead. Loops are initially torsionally relaxed (i.e., linking number, Lk, $=$0). (ii) Pyramids are less well defined
than in (B) and (C), but nevertheless tend to coincide with loops (see also Fig. S9).
(E) Supercoiled loops. (i) As (D), but each loop has a linking number of +32. (ii) Loops form pyramids that are more distinct than in (D).}
\end{figure}

In a homogeneous fiber, pyramids disappear on averaging because domains form stochastically; however, if the fiber is suitably patterned, domain boundaries form at specific locations. For example, if long blocks in which every 20$^{\rm th}$ bead is pink (binding red factors) are interleaved with short blocks containing non-binding grey beads (representing gene-poor ``deserts''), pink beads cluster but grey ones do not; then, many contacts are seen between pink beads to give pyramids sitting exactly on long segments ({Fig. 3B}). Here, boundaries between domains are located within grey segments. Domains are also seen if segments containing 300 successive pink beads (binding red factors) are interleaved with shorter segments containing 100 light-green beads (binding green factors); in this case, a repeating pattern of large and small pyramids is seen, with boundaries between blocks of differently-colored beads ({Fig. 3C}). This simulation could mimic binding of polymerizing complexes and HP1$\alpha$ (hetero-chromatin protein 1$\alpha$) to repeats of eu- and hetero-chromatin~\cite{Kilic2015}. [These results confirm and extend those obtained using Monte Carlo simulations of just one segment of each type~\cite{Barbieri2012,Hofmann2015}.] These runs of binding beads give larger clusters (i.e., $\sim$40 red factors/cluster, and $\sim$15 green factors/cluster). When the fiber is forced permanently into loops (perhaps maintained by CTCF) -- and if red factors can bind to any bead -- pyramids (which are more blurred than in {Fig. 3B} and { C}) tend to sit over each loop ({Fig. 3D}~\cite{Le2013}); this is reminiscent of Hi-C data~\cite{Dixon2012,Le2013,Rao2014}. Therefore, domains can form in a non-uniform genomic landscape ({Fig. 3B}, { C}), and if there are loops ({Fig. 3D}). If loops are preformed into left-handed (or right-handed) inter-wound supercoils~\cite{Gilbert2014}, pyramids are more sharply defined ({Fig. 3E}; see also~\cite{Le2013,Benedetti2014}).

Because our fibers form many ($>$10) domains, we can analyze contact maps away from the diagonal: these clearly show that in all cases where domains form, inter-domain interactions are weaker than intra-domain ones (compare Fig. { S7B}-{ E}) – as in Hi-C data. [Note that many domain:domain interactions seen after one run ({Fig. S7F}) disappear (or become fainter) after averaging data from many runs ({Fig. S7B}).]

\subsection*{Some characteristics of domains}

The probability that two loci yield a Hi-C contact decreases as the number of intervening base pairs increases~\cite{Lieberman-Aiden2009}, and the exponent ($\alpha$) in the power law varies from -0.5 (in HeLa~\cite{Naumova2013}) to -1.6 (in embryonic stem cells~\cite{Barbieri2012}). [$\alpha$ = 1 in the fractal globule model~\cite{Lieberman-Aiden2009,Naumova2013}.] In all simulations in {Figure 3} (except for {Fig. 3A}), there are two regimes -- below and above the domain size (the largest of the two domains appears to set the scale) with $\alpha$ between -0.6 and -1 (strong interactions within a pyramid/domain) or close to -2 (weaker interactions between pyramids/domains; {Fig. S8}). Therefore our values are similar to those seen experimentally. Our results are also consistent with the exponent (seen in simulations of uniform fibers) varying with protein number and affinity~\cite{Barbieri2012}.

We next used various approaches to identify domain boundaries ({Fig. S9}). Many current approaches are based on what we will call a Janus plot ({Fig. S9}), which in its simplest form quantifies all contacts that one bead makes with others to the right or the left in 1D genomic space. In {Figure 3E}, peaks in the two resulting plots correlate well with the left and right tethers of a loop ({Fig. S9A}). By subtracting signal from the two plots, we obtain a ``difference plot'' (i.e., the number of contacts to the right minus the number of contacts to the left). At a boundary, we expect a bead to switch its contacts, from mostly leftward to mostly rightward; consequently, boundaries are found at points where signal in the difference plot crosses zero with an upward derivative ({Fig. S9B}). This is essentially the method used in~\cite{Dixon2012}. In the case of {Figure 3E}, this approach finds domains within loops, and boundaries somewhere in the linear region between them. A more accurate determination is possible by locating the peaks of the derivative of the difference plot (the ``insulator'' plot in {Fig. S9C}). This peak-finding algorithm is elegant and works well with highly-sampled contact maps (as in {Fig. 3}). However, it works less well with sparser data from simulations and Hi-C, where we found the ``difference'' plot gave better results (so we use it in what follows, aided by visual inspection to fine-tune boundary positions).

\subsection*{Modeling selected regions of the human genome}

Finally, we examined whether binding of just two proteins to ``active'' and ``inactive'' beads on a 15-Mbp fiber (representing part of chromosome 12 in GM12878) could fold the genome appropriately ({Fig. 4A}; { Movies S5} and { S6}). Active regions were selected using the Broad ChromHMM track on the UCSC (University of California at Santa Cruz) browser~\cite{Ernst2011}, and beads (1 kbp) representing active promoters or strong enhancers (states 1, 4, 5) -- and the bodies of active transcription units (states 9, 10) -- were colored pink and light-green, respectively. These pink and light-green beads bind red factors (transcription factors, polymerizing complexes) with high and low affinities. Inactive heterochromatin was represented by grey beads that bind black proteins (e.g., dimers of HP1$\alpha$~\cite{Kilic2015}). Heterochromatic beads were identified as those having a low GC content -- an excellent and flexible marker (in principle, choice of threshold can allow any fraction of the region of interest to be classified as heterochromatin). Here, $<$41.8\% GC was chosen as the threshold, as this led to the same percentage of heterochromatin in the 15 Mbp as that marked by state 13 (generic heterochromatin) in the HMM track. Other beads (blue) were non-binding. As before, distinct clusters of bound red or black proteins develop with $\sim$14 or $\sim$190 proteins/cluster, respectively (long runs of adjacent grey beads form the larger clusters; {Fig. 4B}). The resulting contact map was strikingly similar to the Hi-C one~\cite{Rao2014}, with simulations correctly predicting 75\% of the Hi-C domain boundaries to within 100 kbp ({Figs 4C}, { D} and { S10A}, { B}; { Table S2}). Boundaries were in this case found using the difference plot aided by visual inspection ({Fig. S10}); purely automated detection correctly locates $\sim$59\% Hi-C boundaries within 100 kbp, which is still statistically significant (this is an underestimate due to algorithmic errors, some examples of which are noted in {Fig. S10}). Simulations also yield a more-ordered rosettogram than any seen previously ({Fig. S10C}); this is consistent with evolution selecting for a genetic and epigenetic sequence that gives ordered rosettes. Similar results were obtained with a 15-Mbp segment of a different chromosome (chr6) in a different cell type ({Fig. S11}). [Here, heterochromatic regions were defined using either \%GC or HMM state 13, but fewer boundaries were correctly reproduced using the latter ({Fig. S11E}).]

\begin{figure}
\centerline{\includegraphics[width=10.cm]{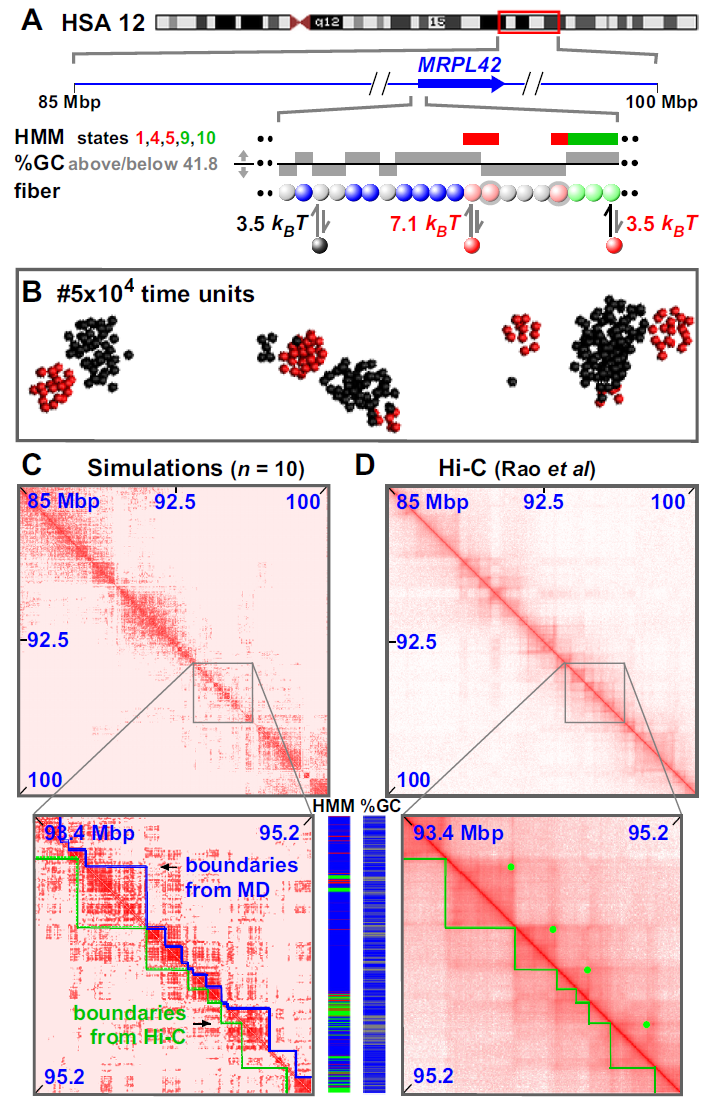}}
\caption{
{\bf Simulating 15 Mbp of chromosome 12 in GM12878 cells.} 
Conditions as Figure 1, with exceptions indicated (chromatin concentration now ~0.01\%).
(A) Overview. The ideogram (red box gives region analyzed) and Broad ChromHMM track (colored regions reflect chromatin states) are from the UCSC browser; the zoom illustrates the MRPL42 promoter. Beads (1 kbp) are colored according to HMM state and GC content (blue -- non-binding; pink -- states 1+4+5, $n=$600; light-green – states 9+10, $n=$880; grey $<$41.8\% GC, $n=$10,646). Red factors (n = 300) bind to (active) pink and light-green beads with high and low affinities, respectively; black (heterochromatin-binding) proteins ($n=$3,000) bind to grey beads.
In the zoom, two pink beads (grey halos) bind both red factors and black proteins.
(B) Snapshot (without chromatin) of central region after 5$\times$10$^4$ time units; most clusters contain factors/proteins of one color. Long runs of grey beads form large black clusters.
(C,D) Contact maps from simulations (7 kbp binning) and Hi-C (10 kbp binning; Rao et al., 2014). In zooms, blue and green lines mark boundaries determined by visual inspection of data
from simulations or Hi-C, and dots in D mark loops found using the Janus plot (Fig. S9A).
Tracks of HMM state and \%GC (colored as in A) illustrate correlations with domains and boundaries.}
\end{figure}

These successes prompted us to model a whole 59-Mbp chromosome (chr19; see Fig. 5 and { Movie S7}). Active and inactive beads (each now representing 3 kbp) were defined as before, and heterochromatic ones using $<$48.4\% GC (to reproduce the fraction of the chromosome bearing the state 13 mark). Now, 85\% domain boundaries are correctly reproduced to within 100 kbp ({Figs 5}, { S12A}). Moreover, simulation boundaries are rich in ``active'' and non-binding regions and depleted of ``inactive'' ones ({Fig. 12B}). As before, the rosettogram and $f_d$ point to a highly-ordered structure with many local contacts ({Fig. S12C}). At a higher level in the organization, 3D positioning of some domains next to others – reflected by off-diagonal blocks in contact maps -- is sometimes reproduced in simulations (zooms in {Figs 5C} and { D}, { S12A}). Our simulations of the whole chromosome further indicate that small active domains seem to be often located at the periphery of larger inactive ones; they also suggest that active domains are more dynamic and mobile than inactive ones (see { Movie S7}). 

\centerline{\includegraphics[width=10.cm]{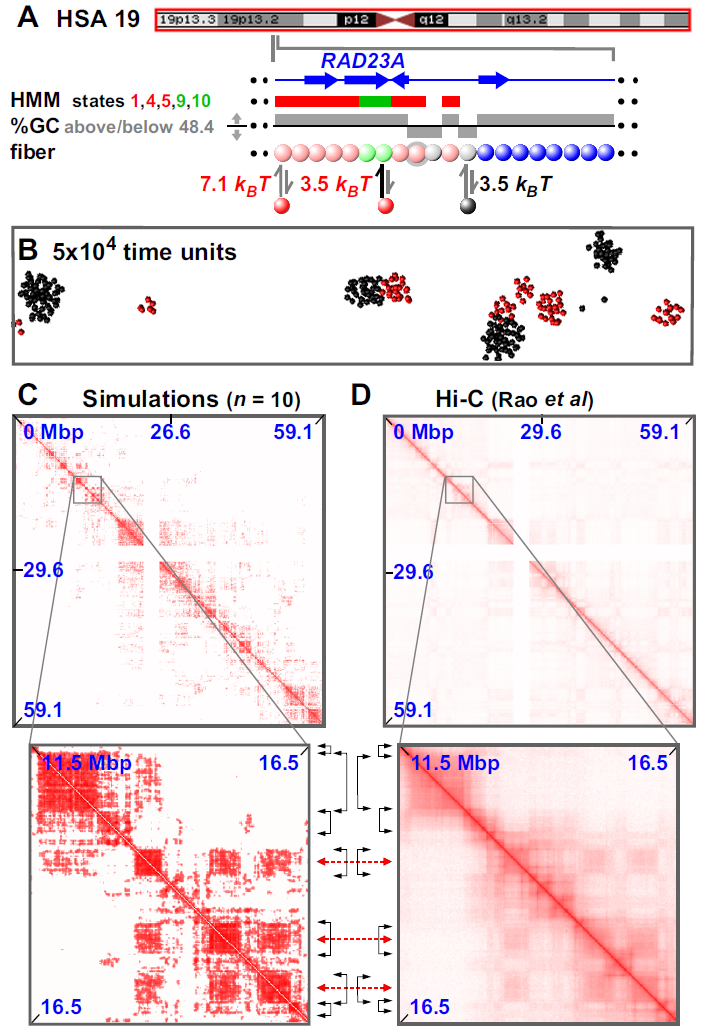}}
\captionof{figure}{\label{Fig5} 
Simulating chromosome 19 in GM12878 cells. Conditions as Figure 4, with exceptions indicated.
(A) Overview. The ideogram (red box indicates whole chromosome simulated) and HMM track (colored regions reflect chromatin states) are from the UCSC browser; the zoom illustrates the region around RAD23A. Beads (3 kbp) are colored according to HMM state and GC content (blue -- non-binding; pink -- states 1+4+5, $n=$2,473; light-green -- states 9+10, $n=$2,686; grey $<$48.4\% GC,
$n=$9,472). Red factors ($n=$400) bind to (active) pink and light-green beads with high and low affinities, respectively; black (heterochromatin-binding) proteins (n = 4,000) bind to grey beads. In the zoom, two pink beads (grey halos) bind both red and black factors.
(B) Snapshot (without chromatin) of central region after 5$\times$10$^4$ units; most clusters contain factors (or proteins) of one color.
(C,D) Contact maps (21 and 20 kbp binning for data from simulations and Hi-C). Between zooms, black double-headed arrows mark boundaries of prominent domains (on the diagonal), and red double-headed ones the centers of off-diagonal blocks making many inter-domain contacts (boundaries and domains detected via the difference plot aided by visual inspection).}

\section*{Discussion}

These MD simulations illustrate some emergent properties of a minimalist system that involves bivalent ``transcription factors'' (or ``proteins'') binding specifically and transiently to cognate sites in a fiber (representing ``chromatin''). First, bound factors spontaneously cluster -- to compact the fibers ({Figs 1}, { 2C}). This self-organization occurs in the absence of any explicit interaction between factors or between beads, and it is driven by a combination of forces dubbed the ``bridging-induced attraction''. Second, and more surprisingly, factors binding to distinct sites on the fiber self-assemble into distinct (segregated) clusters. For example, bound red and green factors self-assemble into clusters that contain only red factors, or only green ones -- but rarely both ({Fig. 2A}). These clusters arise because protein binding will inevitably yield clusters in different places in 3D space if -- and only if -- their cognate binding sites are spatially separated in 1D sequence space ({Fig. 2B}). Third, clustering organizes the loops caused by binding into higher-order rosettes, and domains ({Figs 3}-{ 5}). For example, binding of just two ``proteins'' (transcription and HP1$\alpha$ complexes) to active and inactive regions in a 59-Mbp human chromosome folds the fiber to yield a contact map in which $\sim$85\% of the Hi-C boundaries are both correctly placed and rich in the appropriate sequences ({Figs 5} and { S12}, { Movie S7}). In other words, complexes bind locally to create loops, bound complexes cluster together with similar ones into rosettes, this folds the fiber globally into appropriate domains, and domains pack against each other -- all in the expected ways. Remarkably, then, this minimalist system generates structures that possess all the key features of interphase chromosomes outlined in the Introduction. Moreover, the clusters formed are reminiscent of nuclear structures like Cajal and promyelocytic leukemia bodies, which are each rich in distinct proteins that bind to different cognate DNA sequences~\cite{Sleeman2014}; they also closely resemble nucleoplasmic transcription factories that each contain $\sim$10 active polymerizing complexes~\cite{Pombo1999,Cook1999,Papantonis2013}.

The binding energy of any one factor is small (roughly comparable to that in a few H-bonds), but extended genomic regions fold simply because so many are involved. Cluster formation and protein-driven chromosome organization also occurs quickly (within minutes according to our simulations, {Fig. 1E}). Once clusters form, they usually persist ({Fig. 1E}). However, the system can evolve when new factors appear ({Fig. S6}), much as a transcription factory can develop into a replication factory at the beginning of S phase~\cite{Hassan1994}, or into one specializing in transcribing responsive genes during the inflammatory response (when tumor necrosis factor $\alpha$ induces nuclear influx of nuclear factor $\kappa$B~\cite{Papantonis2012}.

Contacts made as a result of such clustering involve sites both near and far apart on the fiber. Most contacts are local, to create loops and ``rosettes'' like those often seen in models of chromosomes ({Figs 1F}, { 2Aiv}, see also~\cite{Pienta1984,Cook1995}). We suggest rosettes (with many transitive loops) are favoured over more disordered non-local structures (with many overlapping loops) partly because the entropic cost is less~\cite{Marenduzzo2009}; rosettes are also likely to be kinetically favoured when starting from knot-free structures (both in simulations and on exit from mitosis). More distant contacts also depend on patterns of binding sites in 1D genomic space, as domains separated by well-defined boundaries result if non-binding gene deserts or supercoiled loops~\cite{Benedetti2014} are introduced, or if blocks of eu- and hetero-chromatin alternate ({Fig. 3}).
 
In summary, the bridging-induced attraction provides a robust, simple, and generic mechanism that can concentrate specific proteins bound to cognate sites into clusters, and fold interphase fibers in ways found in vivo. Then, the system must either spend energy to prevent the resulting clustering, or -- as seems likely -- it goes with the flow and uses other more or less familiar forces (charge interactions, H bonds, van der Waals, hydrophobic forces, and the depletion attraction~\cite{Marenduzzo2006}) to organize those clusters. If so, we suggest that the particular folding pattern found in any one nucleus will be largely determined by which transcription factors bind to cognate sites, and which bound factors then happen to co-cluster. We also expect that adding more proteins and fibers to our simple model will improve the concordance between contact maps obtained from simulations and Hi-C.

After the present work was completed, two studies proposed another model for formation of looping domains that is based on CTCF bridges~\cite{Mirny2015,LibAid2015}. These involve some loop-extruding driving domain formation, and they are appealing because they can account for the recent observation that CTCF bridging depends on the orientation of the cognate binding sites~\cite{Rao2014,CRISPCTCF}. However, this model requires some as-yet undiscovered motor protein with a processivity sufficient to generate loops of hundreds of kb. On the other hand, as discussed in Ref.~\cite{LibAid2015}, these studies do not address what might underlie the observed compartmentalization into active and inactive domains - which is naturally explained within our framework by binding of different factors to eu- and hetero-chromatin. Furthermore, knock-outs of CTCF have only minor effects on domain organization~\cite{Zuin2014,Hou2010}, which again suggests that this factor cannot be the sole organizer. The results of these knock-outs are also naturally explained by our model, as the compartmentalization is driven by bivalent factors which are unrelated to CTCF. Our work and those of Refs.~\cite{Mirny2015,LibAid2015} are therefore complementary, and it would be of interest to couple the two approaches together in the future. 

\section*{Materials and Methods}

Brownian dynamics (BD) simulations were run with the LAMMPS (Large-scale Atomic/Molecular Massively Parallel Simulator) code, used with a stochastic thermostat. Chromatin fibers are modeled as bead-and-spring polymers using FENE bonds (maximum extension 1.6 times bead diameter, $\sigma$) and a bending potential that allows persistence length to be set. Protein:protein and chromatin:chromatin interactions involve only steric repulsion, and those between proteins and their binding sites are truncated and shifted Lennard-Jones interactions. All beads are confined within a cube with periodic boundary conditions. Additional details are listed in Figure legends and/or Supplementary Information.

Simulations of human chromosomes included two kinds of factors/proteins, one binding to active euchromatin and the other to inactive heterochromatin. Chromatin beads were colored pink or light-green if interacting strongly or weakly, respectively, with red factors (representing factors and polymerases), grey if interacting with black proteins (representing HP1$\alpha$), and blue if non-interacting. The Broad ChromHMM track~\cite{Ernst2011} on the hg19 assembly of the UCSC Genome browser was used to determine pink/light green color: (i) if a region of 90 bp or more within one bead is labeled as an ``Active Promoter'' or ``Strong Enhancer'' (states 1,4,5) then the whole bead is colored pink, similarly if a bead contains 90 bp or more of ``Transcriptional Transition'' or ``Transcriptional Elongation'' (states 9,10) it is colored light green. Data for GC content ({Figs 4}, {5} and {S11D}) or from the HMM track ({Fig. S11A}, {B}) were used to determine grey color: (i) if GC content of the bead was below the threshold specified in each Figure legend, or (ii) if a region of 90 bp or more within one bead is labelled ``Heterochromatin; low signal'' (state 13). This convention allows one bead to have multiple colors. Figure legends give numbers of differently-colored beads in a simulation and affinities. Simulations  included red factors ($n$=300 in {Fig. 4} and {Fig. S11}; $n$=400 in {Fig. 5}), and black protein ($n$=3,000 in Fig. 4 and Fig. S11; $n$=4,000 in {Fig. 5}); for simplicity, protein size is equal to $\sigma$, and interaction range to 1.8 $\sigma$.

In our simulations, one bead corresponds either to 3 ({Figs 1}-{3}, {5}) or 1 kbp ({Figs 4} and {S11}). These correspond to a size of 30 nm and 20.8 nm respectively (assuming a 30-nm fiber has a packing density of 1 kbp/10 nm, and the two bead sizes contain the same DNA density). Persistence length was 3 $\sigma$ (representing a flexible polymer), and one time unit corresponds to 0.6 or 0.2 ms (for 3 or 1 kbp beads; calculated assuming a viscosity of 10 cP, or 10-fold larger than water).

CAB and DM acknowledge support through an ERC Consolidator Grant (THREEDCELLPHYSICS, Ref.648050).

\newpage

\appendix{Supplementary material}

\section{Coarse grained molecular dynamics simulations}

In our coarse grained molecular dynamics simulations, we represented chromatin as a bead-and-spring chain, and protein complexes as additional beads. The position of the $i$th bead in the system changes in time according to the Langevin equation
\begin{equation}\label{langevin}
m_i \frac{ d^2 \mathbf{r}_i }{dt^2} = -\nabla U_i - \gamma_i \frac{d\mathbf{r}_i }{dt} + \sqrt{2k_BT\gamma_i}\boldsymbol{\eta}_i(t),
\end{equation}
where $\mathbf{r}_i$ is the position of bead $i$, $m_i$ is its mass, $\gamma_i$ is the friction it feels due to an implicit aqueous solvent, while $\boldsymbol{\eta}_i$ is a vector representing random uncorrelated noise which obeys the following relations
\begin{equation}
  \langle \eta_{\alpha}(t) \rangle = 0 ~~\mbox{and}~~  \langle \eta_{\alpha}(t)\eta_{\beta}(t') \rangle = \delta_{\alpha\beta} \delta(t-t').
\end{equation}
The noise is scaled by the energy of the system, given by the Boltzmann factor $k_B$ multiplied by the temperature of the system $T$, taken to be 310~K for a cell. The potential $U_i$ is a sum of interactions between bead $i$ and all other beads, and we use phenomenological force fields as described below. For simplicity we assume that all beads in the system have the same mass and friction $m_i\equiv m$, and $\gamma_i\equiv \gamma$. Eq.~(\ref{langevin}) is solved in LAMMPS using a standard Velocity-Verlet algorithm.

\section{Force fields}\label{forcefield}

For the chromatin fiber the $i$th bead in the chain is connected to the $i+1$th with a finitely extensible non-linear elastic (FENE) spring: the associated potential is given by
\begin{align}\label{FENE}
U_{\rm FENE}&(r_{i,i+1}) =& \nonumber\\ &U_{\rm WCA}(r_{i,i+1}) - 
\frac{K_{\rm FENE} R_0^2}{2} \log \left[ 1 - \left(\frac{r_{i,i+1}}{R_0}\right)^2 \right] ,
\end{align}
where $r_{i,i+1}=|\mathbf{r}_i-\mathbf{r}_{i+1}|$ is the separation of the beads, and the first term is the Weeks-Chandler-Andersen (WCA) potential
\begin{align}
\frac{U_{\rm WCA}(r_{ij})}{k_BT}  = \left\{ 
\begin{array}{ll} 
4 \left[ \left( \frac{d_{ij}}{r_{ij}}\right)^{12} - \left( \frac{d_{ij}}{r_{ij}}\right)^{6} \right] + 1, & r_{ij}<2^{1/6}d_{ij} \\
0, & \mbox{otherwise},
\end{array} \right.
\label{eq:WCA}
\end{align}
which represents a hard sphere-like steric interaction preventing adjacent beads from overlapping. In Eq.~(\ref{eq:WCA}) $d_{ij}$ is the mean of the diameters of beads $i$ and $j$. The diameter of the chromatin beads is a natural length scale with which to parametrize the system; we denote this by $\sigma$, and use this to measure all other length scales.  The second term in Eq.~(\ref{FENE}) gives the maximum extension of the bond, $R_0$; throughout this work we use $R_0=1.6~\sigma$, and set the bond energy $K_{\rm FENE}=30~k_BT$. 

The bending rigidity of the polymer is introduced via a Kratky-Porod potential for every three adjacent DNA beads
\begin{equation}\label{eq:bend}
U_{\rm BEND}(\theta)=\\K_{\rm BEND} \left[ 1 - \cos(\theta) \right],
\end{equation} 
where $\theta$ is the angle between the three beads as give by
\begin{equation}
\cos(\theta) = [\mathbf{r}_i-\mathbf{r}_{i-1}]\cdot [\mathbf{r}_{i+1}-\mathbf{r}_{i}],
\end{equation}
and $K_{\rm BEND}$ is the bending energy.  The persistence length in units of $\sigma$ is given by $l_p=K_{\rm BEND}/k_BT$. 

Finally, steric interactions between non-adjacent DNA beads are also given by the WCA potential [Eq.~(\ref{eq:WCA})]. In the absence of proteins, the force field of chromatin is therefore appropriate for a biopolymer in a good solvent. 

Each protein (or protein complex) we simulate is represented by a single bead; unless otherwise stated, the WCA potential is used to model steric interactions between these. Chromatin beads are labeled as binding or not-binding for each protein species according to the input data. For the interaction between proteins and the chromatin beads labeled as binding, we use a shifted, truncated  Lennard-Jones potential, whose form is given by
\begin{equation}
U_{\rm LJcut}(r_{ij})= \left\{ \begin{array}{cl} 
U_{\rm LJ0}(r_{ij}) - U_{\rm LJ0}(r_{\rm cut}) & r_{ij}<r_{\rm cut}, \\0 & \mbox{otherwise}, \end{array}\right. 
\label{eq:LJ}
\end{equation}
with
\[
U_{\rm LJ0}(r)=4\epsilon' \left[ \left( \frac{d_{ij}}{r}\right)^{12} -  \left( \frac{d_{ij}}{r}\right)^{6} \right],
\]
where $r_{\rm cut}$ is a cut off distance, and $r_{ij}$ and $d_{ij}$ are the separation and mean diameter of the two beads respectively. This leads to an attraction between a protein and a chromatin bead if their centres are within a distance $r_{\rm cut}$. Here $\epsilon'$ is an energy scale, but due to the second term in Eq.~(\ref{eq:LJ}) this is not the same as the minimum of the potential, which for clarity we denote as $\epsilon$ (and we refer this to as the interaction energy). For simplicity we set the diameter of the protein complexes equal to that of the chromatin beads, $d_{ij}=\sigma$, and set $r_{\rm cut}=1.3~\sigma$ unless otherwise stated.

The length scale $\sigma$, mass $m$ and energy scale $k_BT$ give rise to a natural simulation time unit $\tau_{\rm LJ}=\sqrt{\sigma^2 m /k_BT}$, and Eq.~(\ref{langevin}) is integrated with a constant time step $\Delta t=0.01\tau_{\rm LJ}$, for a total of $6\times10^6$ time steps or more (see main text).

\section{Mapping simulation units to physical units}

In order to compare simulation and experimental time and length scales, it is useful here to describe how to map simulation into physical units (this is not required for energy as this was previously expressed in units of $k_BT$). 

Length scales are easily mapped once the value of $\sigma$ is set in physical units. For simulations of chromatin fibers where one bead corresponds to 3 kbp, a natural choice is $\sigma=$30 nm, leading to a linear baseline packing of 10 nm/kbp. For the higher resolution simulations of the chr12 and chr6 regions ({ Figs 4} and { S11}), $\sigma$ corresponds to 1~kbp. Assuming the same chromatin density in the two models, a unit of length now corresponds to 20.8 nm.

In order to map time units, we need to recognise that there are three main time scales in the system. One is the previously defined Lennard-Jones time $\tau_{LJ}$. A second is the inertial time $\tau_{\rm in}=m/\gamma_i$ (from Eq.~(\ref{langevin})), which is the characteristic time over which a bead loses information about its velocity. A third typical time is the so-called Brownian time $\tau_{\rm B}=\sigma^2/D_i$, which gives the order of magnitude of the time it takes for a bead to diffuse across its own diameter $\sigma$. Here $D_i$ is the diffusion constant for bead $i$, given through the Einstein relation by $D_i=k_BT / \gamma_i$. If we make the approximation that a chromatin bead will diffuse like a sphere we can then use Stokes' law, where $\gamma_i=3\pi\nu d_i$, with $\nu$ the viscosity of the fluid, and $d_i$ the diameter of bead $i$. Taking realistic values for the length, mass and viscosity one finds that $\tau_{\rm in} \ll \tau_{\rm LJ} \ll \tau_{\rm B}$, with the times separated by several orders of magnitude. For numerical stability we must choose the time step $\Delta t$ smaller than all of these times, and we wish to study phenomena which will occur on times of the order $\tau_{\rm B}$; this means that using real values for all parameters would lead to unfeasibly long run times. Instead we chose parameters such that $\tau_{\rm in} \le \tau_{\rm LJ} \le \tau_{\rm B}$, and map from simulation to physical time scales through the Brownian time $\tau_{\rm B}$. This assumption means that processes which occur on time-scales below the Brownian time are not resolved accurately, however this is of no practical consequence for our work as we are interested in time-scales much exceeding the Brownian time. 

For simulations where chromatin beads were 30 nm in diameter (all except { Figs 4} and { S11}), taking a viscosity of 10 cP for the nucleoplasm (10 times that of water, to account for the effective increase in viscosity due to crowding) gives a Brownian time of about 0.6 ms, so that a simulation run of $5\times 10^6$ time steps corresponds to about 30 s of real time. For simulations where chromatin beads correspond to 1 kbp (20.8 nm in diameter), one simulation unit of time (one Brownian time) corresponds to about 0.2 ms.

\section{Additional simulation details}

For the simulation in { Figures 3E} and { 3F}, the force field discussed in Section~\ref{forcefield} was supplemented with
torsional interactions to generate results for loops with linking number Lk
equal to 0 or 32. To model supercoiled or torsionally
relaxed (but not nicked) loops, we use closed loops (each of contour
length 324 $\sigma$), which were joined to a linear backbone with a
Gaussian spring. We modeled torsional interactions using spherical atoms
with an associated triad of vectors, so that the Euler angles describing the
relative orientation of adjacent beads allow us to track the twist as well as
the bending rigidity. This scheme corresponds to model 2 described 
in Ref.~\cite{Brackley2013}; chromatin was modeled as a ribbon in the torsionally
relaxed state, and with torsional persistence length equal to 20 $\sigma$.

For convenience, we also list here Lennard-Jones parameters for all attractive interactions in the simulations (the rest of the interactions are repulsive and modeled using a WCA potential, as previously mentioned). The interaction range (cut-off of Lennard-Jones interaction) was equal to 1.8 $\sigma$ for all attractive interactions. Interaction strengths ($\epsilon'$, in units of $k_BT$) were as follows: { Figures 1}, { 3A}, { 3B} and { S2B}: 7.1 (between red ``transcription factors'' and pink beads); 3.5 (between red factors and blue beads). { Figure 2A}: 8.9 (between red factors and pink beads; and between green factors and light-green beads). { Figures 2C} and { S5}: 7.1 (between each of the factors and its target binding beads). { Figures 3C}: 3.5 (between red factors and pink beads; and between green factors and light-green beads). { Figures 3D} and { 3E}: 3.5 (between red factors and pink beads. { Figures 4}, { 5} and { S11}): 7.1 (between red factors and pink beads); 3.5 (between red factors and light-green beads; and between black ``proteins'' and grey beads). { Figure S2A:} 7.1 (between red factors and pink beads). { Figure S3}: 7.1 (between either red or green factors and yellow beads). { Figure S4}: 7.1  (between red factors and pink beads; and between green factors and light-green beads). { Figure S6}: initially 7.1 only between red factors and pink beads; after ``switch'' 7.1 (between red factors and pink beads), and 13.1 (between green factors and light-green beads).

\section{Initialization}

Finally, as in all molecular dynamics simulations it is important to specify how the system was initialised. For all cases where a single linear polymer was modeled, chromatin fibers were first generated as random walks, and proteins randomly distributed (with uniform probability throughout the simulation box). The simulation was then run with a soft potential between all beads to remove overlaps, and with a Gaussian spring between neighboring beads (this was for at least a million time steps; in some cases it was also necessary to use a higher bending rigidity to avoid initial entanglements). After equilibration, the force field was set to the one discussed in Section~\ref{forcefield}. For { Figures 3E} and { 3F}, we first equilibrated supercoiled or torsionally relaxed loops in isolation, then joined them to a linear backbone at appropriate places (see caption to { Fig. 3}) with Gaussian springs; the system was then allowed to equilibrate with the force field in Section~\ref{forcefield} (which preserves topology and linking number as it disallows intrachain crossings). For { Figure S5}, we generated initial conformations for the 20 chromatin fibers as mitotic-like cylinders with random orientation, following the method described in Ref.~\cite{Rosa2008}; proteins were still distributed randomly and uniformly at the beginning of the simulations. The equilibration steps were then performed as above (with soft potential and Gaussian springs).

\section{Analysing contacts: contact maps, boundaries and rosettograms}

An important output of both Hi-C experiments and our simulations are contact maps; in this Section we discuss how we analysed them.

The contact maps in { Figures 1D}, { 2Aiii}, { S2Aiii}, { S2Biii}, { S3iii}, { S4iii}, { S5iii} and { S7F} were obtained from a single configuration: a contact between two beads was scored if their centers were less than 150 nm (5 $\sigma$) apart. We binned contacts by dividing the polymer into a number of bins (specified in each Figure Legends) to aid visualization. 
The colored contact maps in { Figures 2Aiii}, { S3iii}, { S4iii}, { S5iii} and { S6iv} were also obtained from a single configuration, by only considering the binding sites. A contact between two binding sites was scored if their centers were less than 90 nm (3 $\sigma$) apart. Binding sites were colored according to the protein (or factor) which is attracted to them (i.e., pink sites are colored red as they bind red factors, etc.); in case a binding site could be the target for more than one factor (e.g., in { Fig. S6iv}), we colored the binding site according to the protein which was closest to them (e.g., red if the binding site was closest to a red protein, etc.). Pixels in the contact map then are colored red if they are contacts between two red pixels, etc.; mixed contacts are colored yellow if between red and green, and grey in { Figure S5iii}.
Finally, the simulation contact maps in { Figures 3}, { 4}, { 5}, { S7A-E} and { S11}, were averaged over several realizations (specified in the Figure Legends, together with the binning used).
In all contacts map (with the exception of colored contact maps), the entry gives the number of contacts in the bin, scaled by the maximum number of contact maps over all bins (in this way entries are between 0 and 1). 

For each of the simulations in { Figure 3},  we plotted both the whole contact map ({ Fig. S7}) and just the part of it close to the diagonal (referred to as pyramid plots in the text); the latter is often used in the literature as it allows a clearer visual determination of boundaries. While the simulated contact maps are shown without any normalization, experimental contact maps for GM12878 cells ({ Figs 3} and { 5} in the main text) were normalized according to the square root normalization method described in Ref.~\cite{Rao2014}. Experimental contact maps for { Figure S11} were not normalized; these were computed from the Sequence Read Archives (SRA) data in the Gene Expression Omnibus [obtained from Ref.~\cite{Dixon2012} via access number GSE35156; duplicate reads were removed].

For each of the contact maps (whether from simulation or experiments), we prepared Janus and difference plots, and computed the number of contacts (or contact probability) versus distance along the genome/simulated chromatin fibers. All contact maps were binned (the binning used varied in the different cases and in specified in Figure Legends).

The Janus forward signal for bin $i$, $F(i)$, is defined as the sum over all contact map entries relative to contacts which a bead makes with other beads to the right of it: i.e., $\sum_{j=i+1}^{n} c(i,j)$, where $c(i,j)$ denotes the contact map entry relative to the $i$-th and $j-$th bins, and $n$ is the total number of bins. The Janus backward signal for bin $i$, $B(i)$, was similarly computed as $\sum_{j=1}^{i-1}c(i,j)$. 

The difference plot ({ Figs S8} and { S9}) is the difference $\Delta(i)\equiv F(i)-B(i)$. When this quantity is negative, bin $i$ is making more contact to its left; when it is positive, the majority of the contacts bin $i$ makes are to its right. The difference plot is useful to get a first estimate of domain boundary locations, since boundaries are places where the pattern of contacts made by a bin changes from mostly to the left to mostly to the right (but not vice versa). Therefore, boundaries can be located at regions where $\Delta(i)$ crosses 0 with an upward derivative; a similar algorithm was used to locate boundaries in Ref.~\cite{Dixon2012}. This is the base of the algorithm used in { Figure S9} to determine boundaries automatically in the region chr12:85000000-100000000 bp ({ Fig. 4} in the main text). To avoid spurious multiple nearby boundaries due to noise in the difference plot signal, we further required that either the upward trend in $\Delta(i)$ is common to 4 consecutive beads crossing zero, or that the upward slope at the zero crossing (forward difference $\Delta(i+1)-\Delta(i)$ where $\Delta(i)<0$ and $\Delta(i+1)>0$) is larger than a set threshold (equal to 10\% or 40\% of the maximum step in the function $\Delta(i)$, for experiments and simulations respectively). 

Another way to detect boundaries is via peaks in the derivative of $\Delta(i)$ (this is the insulator plot in { Fig. S8}): the rationale here is that we expect the relative fraction of contacts to the right should increase sharply at boundaries, however due to contacts away from the diagonal it may not necessary be that the difference plot goes through 0. In selected cases we also used the recent method described in Ref.~\cite{Hsieh2015} to detect boundaries. While all methods agreed on some of the boundaries, visual inspection suggests that not all boundaries can be found by any one automated technique (see { Figs 4} and { S10B}). While in Hi-C experiments the numerical error that these or similar algorithms make is not too important, it is much more consequential with simulation data that are noisier. Moreover, our goal is to compare experimental and simulation boundaries, rather than to estimate boundaries in either simulations or experiments with a given accuracy. Comparing boundaries in simulations and Hi-C data is a demanding task: for instance, even a two pixel error (the $\Delta(i)$ curve turning in an opposite direction) would lead to an artificial discrepancy of 80 kbp between the location of the same boundary in simulations and experiments, and missing out boundaries or false detection of boundaries due to noise would give an even more serious reduction in the measure of the agreement between simulations and experiments (as the number of boundaries which can be located randomly is relatively high, see text). As a result, while the automated detection of boundaries in { Figure S9} shows that the agreement between simulations and experiments is statistically significant, there are errors in the boundary detection which affect this comparison. To avoid this, we resorted to locating boundaries by visual inspection (compare { Fig. 4} with { Fig. S9}).

Finally, we discuss some details of rosettograms ({ Fig. S1}). To build these, we start from one configuration from the simulation, and divide the binding beads into clusters; two binding beads are defined to be in the same cluster if their separation is below a threshold (typically 90 nm, unless specified otherwise). Then clusters are numbered, starting from the first along the chromatin fiber ({ Fig. S1}). The rosettogram plots cluster number versus binding bead number, and for clarity we only show the binding beads which are in clusters. A string of well formed rosettes shows up as a series of continuous lines (lines made up by contiguous pixels) in the rosettogram, whereas a more disordered structure with lots of non-local contacts is characterised by breaks in the horizontal lines in the rosettogram (as binding beads in the cluster will often not be contiguous along the polymer chain). To quantify how disordered (i.e., how far from an ideal string of rosettes) the loop network of a chromatin fiber is, we compute the {\it fraction disorganised}, or $f_d$. To define $f_d$, we count the number of steps (upwards or downwards) in the rosettogram. In an ideal string of $N$ rosettes there will be $N-1$ steps, so if $N$ denotes cluster number, we subtract $N-1$ from the number of steps: this gives the number of ``errors'', i.e., of non-local loops in the interaction network. The fraction disorganised is then defined as the number of errors per pixel (i.e., the number of errors divided by the number of pixels in the rosettogram). From this definition, it is apparent that a small value of $f_d$ indicates a structure akin to a regular string of rosettes, whereas a large value indicates a disordered structure with many non-local contacts. 

\section{Analysis of bioinformatic data}

Here, we explain how beads were colored using bioinformatic data in the chromosome simulations ({ Fig. 4}, for chr12:85000000-100000000 bp, { Fig. S11} for chr6:5000000-20000000 bp, and { Fig. 5} for the whole of chr19).

Beads in the simulations can interact either with (black) ``proteins'' binding to heterochromatin (when they are colored grey) or with (red) ``transcription factors'' binging to euchromatin (when they are colored pink or light-green according to binding affinity), or with both (indicated in cartoons by the surrounding halo), or with neither (when they are colored blue). 

Data from the Broad ChromHMM track on the hg19 assembly of the 
UCSC Genome browser were used to determine pink/light green coloring as follows: 
(i) if a region of 90 bp or more within one bead (representing 1 kbp, { Figs 4} and { S11}, or 3 kbp, { Fig. 5}) is labeled as an ``Active Promoter'' or ``Strong Enhancer'' (states 1,4,5 on the Broad ChromHMM track), then the whole bead is colored pink (so it binds with high affinity to red factors representing transcription factor/polymerase complexes); (ii) if a region of 90 bp or more within the sequence covered by one bead is labeled ``Transcriptional Transition'' or ``Transcriptional Elongation'' (states 9 and 10), then that bead is colored light green (and binds red factors with low affinity).

To determine whether a bead should be colored grey (i.e., labeled as heterochromatin) we used one of the following two methods. Either we directly used the Broad ChromHMM data (in { Fig. S11}): if 90 bp or more within the bead is classified as state 13, then the whole bead is classified as grey.
Alternatively (in { Figs 4}, { 5} and { S11}) GC content data from the UCSC Genome Browser were used to color beads by setting a threshold GC content percentage and 
coloring beads grey if they fell below this. Here, the rationale behind this is that heterochromatin and gene poor regions are known to correlate with low GC
content (they are rich in AT). The threshold was set, in each case, so as to
end up with the same overall number of heterochromatic beads as one would 
obtain if beads were colored grey according to the  
Broad ChromHMM track. As a result, the \%GC content threshold used was 43.4\% for chr6, 41.8\% for chr12 and 48.4\% for chr19. For chr19, some of the telomeric sequences are missing for hg19; we have assumed these are not binding to black proteins.
We note that our coloring scheme (both when only using the HMM track and when also using GC content) allows a bead to be of more than one color. This is sensible in view of our coarse graining (a single bead can include both a euchromatic and a heterchromatic region), and also become some genomic regions can be targets for competing chromatin-associating proteins.

\newpage

\begin{table}[h!]
\centering
\begin{tabular}{ |c|c|c|c|c|c|c|c| }
 \hline
{\bf Row} &{ \bf Fig.} & {\bf pattern} & {\bf low aff. sites} & {\bf fd} & {\bf n pixels} & {\bf n clusters} & {\bf \% rosettes} \\  
 \hline
 1 & 1F & every 20 & + & 0.13 $\pm$ 0.02 & 173 $\pm$ 2.4 & 21 $\pm$ 0.8 &  0.37$\pm$ 0.04 \\
 2 & S2Aiv & every 20 & - & 0.2 $\pm$ 0.02 & 146 $\pm$ 3.6 & 20 $\pm$ 0.7 &  0.24 $\pm$ 0.03 \\
 3 & S2Biv & random & + & 0.06 $\pm$ 0.002 & 201 $\pm$ 1.4 & 17 $\pm$ 1.0 &  0.5$\pm$ 0.03 \\
 4 & S2Biv   & random 20 & - & 0.12 $\pm$ 0.01 & 193 $\pm$ 3.1 & 20 $\pm$ 0.8 &  0.35$\pm$ 0.07 \\
 5 & 2Aiv & alt. every 20 & - & 0.51 $\pm$ 0.02 & 92 $\pm$ 5.4 & 19 $\pm$ 1.0 &  0.18$\pm$ 0.03 \\
 6 & S4iv & alt. every 20 & - & 0.13 $\pm$ 0.01 & 186 $\pm$ 2.2 & 25 $\pm$ 0.8 &  0.38$\pm$ 0.06 \\
 \hline
\end{tabular}
\caption{Some properties of rosettes found in different simulations ($n$= 5, except for the case in row 4 when $n$= 6). In the table: (i) $f_d$= fraction disorganized; (ii) Number of pixels: number of pixels in rosettogram; (iii) Number of clusters: number of clusters in rosettogram; (iv) alt.=alternating (pink and light green beads); (v) $\% {\rm rosettes}$= percentage of rows with contiguous pixels.}
\end{table}

\begin{table}
\centering
\begin{tabular}{ |c|c|c|c|c|c|c| }
 \hline
{\bf row} & {\bf thr (kbp)} & {\bf p chr12} & {\bf p chr12 (auto)} & {\bf p chr6 (GC)} & {\bf p chr6 (state)} & {\bf p chr19} \\
\hline
1 & 20 & 0.0011 & 0.01 & 0.35 & 0.61 & 2.6e-11 \\
2 & 50 & 0.0024 & 0.003 & 0.017 & 0.14 & $<$1e-11 \\
3 & 100 & 0.000061 & 0.02 & 0.0011 & 0.13 & $<1$e-11 \\
4 & 150 & 0.00048 & 0.37 & 0.000032 & 0.15 & $<1$e-11 \\
5 & 200 & 0.0027 & 0.28 & 0.00015 & 0.045 & 4.6e-11 \\
\hline
\end{tabular}
\caption{Comparison of boundaries seen in data from simulations and Hi-C. Data is from Figures 4 (chr12), 5 (chr6 using either \%GC or HMM state 12 to identify grey beads) and S11 (chr19). Boundaries were identified by visual inspection, except for ‘chr12, automated’ when boundaries were identified in an automated way (see Fig. S10B). In the table: thr=threshold; auto=automated; GC=using \% GC; state=using only chromatin state track.}
\end{table}

\newpage

\centerline{\includegraphics[width=10.cm]{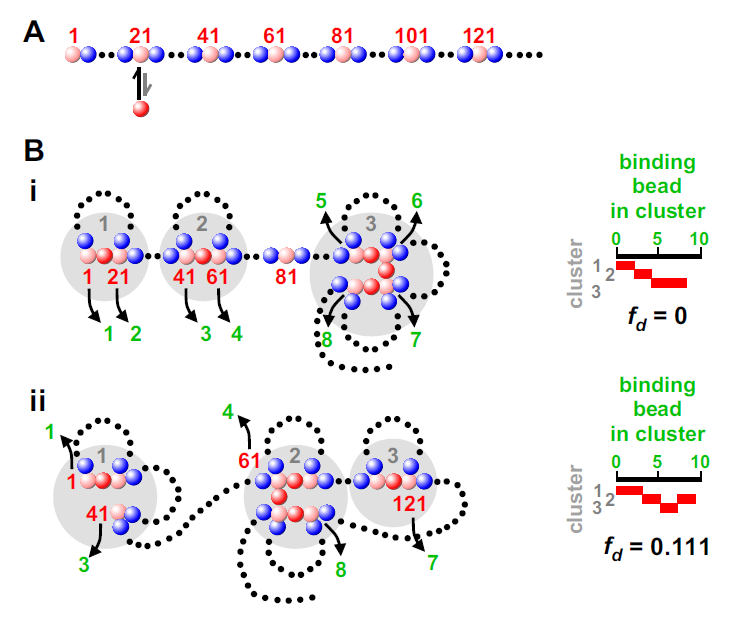}}

\captionof*{figure}
{{\bf Figure S1. Example ``rosettograms''.}
(A) Factors (red) can bind to every 20th bead (pink) in the fiber.
(B) Two possible structures (left), and corresponding rosettograms (right).
(i) A simple structure used to illustrate the numbering system. First, clusters (grey circles) are defined (a cluster contains $\ge$2 binding beads with centers lying $<$90 nm apart). Clusters are then numbered from 1 upwards (grey numbers), beginning with the one containing the lowest- numbered binding bead in the fiber. Binding beads in clusters (but not blue beads, or bead 81 -- which is not in a cluster) are now renumbered as shown in green. In a rosettogram, a red pixel marks the presence of a binding bead in a cluster. In a row, increasing numbers of abutting (conversely, non-abutting) pixels reflect increasing numbers of near neighbor (conversely, distant neighbor) binding beads in a rosette and an organized (conversely, disorganized) structure. The disorganized fraction ($f_d$) is equal to $(S-N+1)/P$, where $S$ is the number of steps in the rosettogram, $N$ is the number of clusters (or rows), and $P$ is the total number of colored pixels (i.e., the total number of binding beads, see Supplementary Information for a motivation for this formula and for a further discussion of fd). This quantity is also equal to the total number of white spaces between first and last colored pixels in each row divided by P (i.e., the total number
of binding beads in clusters). Here, $f_d$ is 0 (the low value reflects an ordered structure where all loops involve nearest-neighbor binding-sites). (ii) A more complicated structure gives a more
complex rosettogram with non-abutting pixels in row 2; as there is one gap between red pixels in row 2 and 9 pixels in all, $f_d = 1/9$.}

\newpage

\centerline{\includegraphics[width=8.cm]{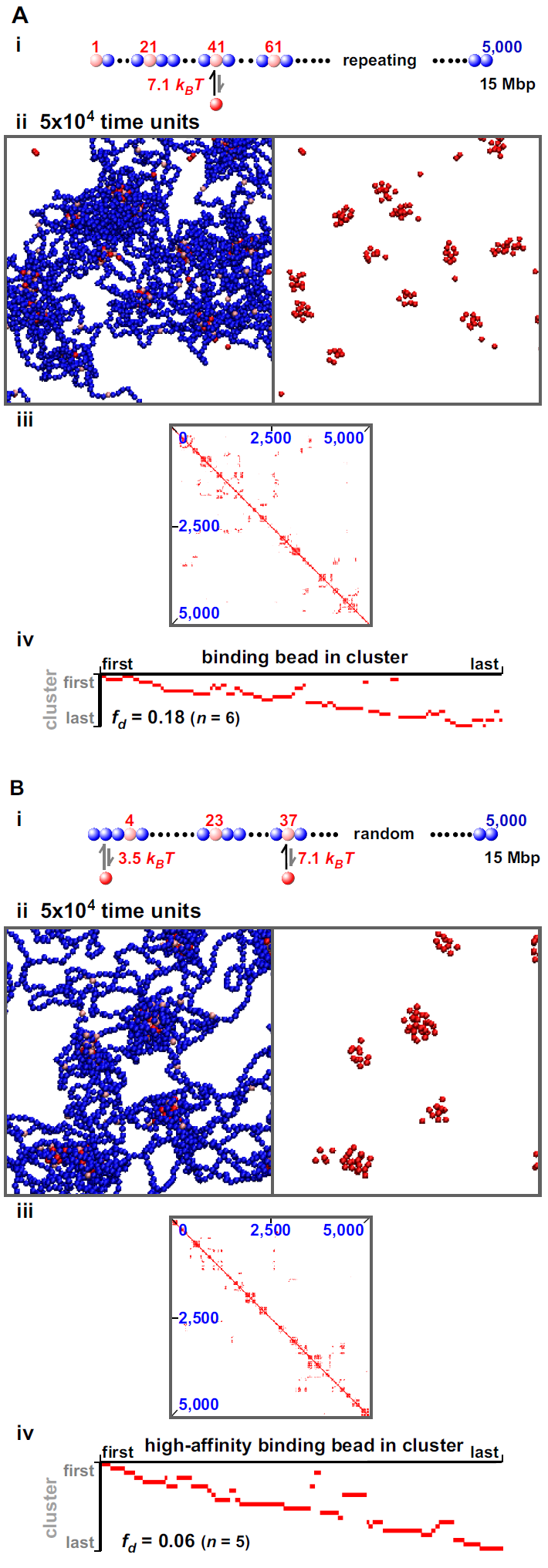}}
\captionof*{figure}
{{\bf Figure S2. Simulations as in Figure 1, showing bound factors still spontaneously cluster if low-affinity binding is absent, or high-affinity sites are randomly distributed.}
(A) Absence of low-affinity binding. (i) Factors have a high affinity for pink beads, but zero affinity for blue beads. (ii) Final snapshots of a central region with/without chromatin; clusters still form.
(iii) Final contact map; blocks along the diagonal are slightly less prominent compared to those seen in Figure 1D (same binning used). (iv) Final rosettogram; most clusters still contain $\ge$2
petals, but runs of abutting pixels in one row are slightly shorter than those seen in Figure 1F (and the $f_d$ is higher, indicating a higher-fraction of non-local loops).
(B) Randomly-distributed binding sites. (i) The same number of pink beads found in Figure 1A are distributed randomly along the fiber. (ii) Final snapshots of a central region with/without chromatin;
clusters still form. (iii) Final contact map; blocks along the diagonal are not so uniform and are spaced irregularly. (iv) Final rosettogram. Perhaps surprisingly, the structure is slightly less disorganized than in Figures 1F and Aiv above (reflected by a lower $f_d$). This is probably because
gaps between successive binding sites are exponentially distributed so that binding sites are naturally clustered nearer together in 1D genomic space (``Poisson clumping''), and this facilitates formation of more ``perfect'' rosettes containing near-neighbor binding beads.}

\newpage

\centerline{\includegraphics[width=10.cm]{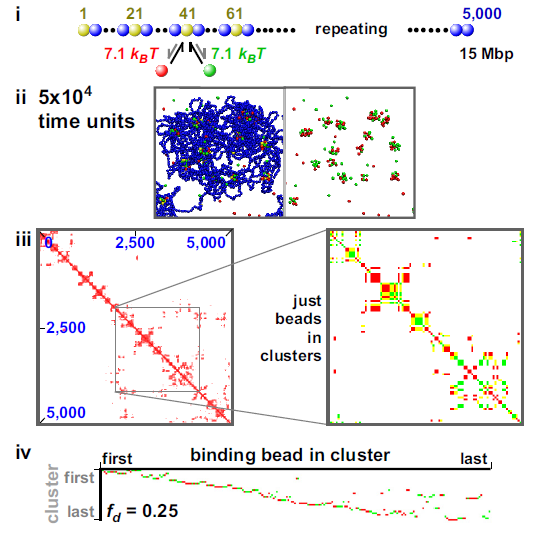}}
\captionof*{figure}
{{\bf Figure S3. Mixed clusters form if red and green factors can bind to the same high-affinity sites.} 
MD simulations were as in Figure 2A, with the differences indicated.
(i) Red ($n$= 250) and green ($n$= 250) factors interact solely with every 20th bead (yellow); data
below are for the state after 5$\times$ 10$^4$ time units. (ii) Final snapshots (± chromatin); bound red and
green factors are often found in one cluster. (iii) Final contact map for all beads (axes give bead
numbers). The inset shows a high-resolution map of just binding beads in clusters (prepared as in
Fig. 2Aiii). Here, red, green and yellow pixels mark contacts between two pink beads (in a
cluster and bound to a red protein), between two light-green beads (in a cluster and bound to a
green protein), and between a light-green and pink bead, respectively. The many yellow pixels
reflect the presence of mixed clusters containing both red and green factors. Note that the patterns of pixels in the regular (left) and high-resolution maps (right) differ slightly both here and in maps shown later; this is the result of the different criteria used to define contacts, and whether binning was used. (iv) Rosettogram (pixels colored according to which high-affinity beads are in the cluster). Rows often contain contiguous pixels of different colors, again reflecting the presence of both types of factor in one cluster. Intriguingly, the fd is higher than that seen with the distinct clusters in Figure S2iv, presumably, this is due to the higher number of proteins in the simulation (500 as opposed to 250).}

\newpage

\centerline{\includegraphics[width=10.cm]{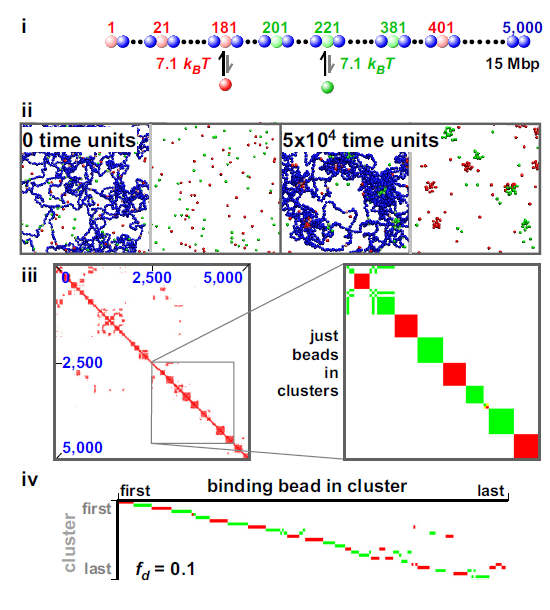}}
\captionof*{figure}
{{\bf Figure S4. Red and green proteins form distinct clusters if their cognate sites are present in
distinct blocks (mimicking eu- and hetero-chromatin).} 
Conditions as in Figure 1, with
exceptions indicated. (i) Red ($n$= 250) and green ($n$= 250) factors interact solely with pink and
light-green beads, respectively; 10 pink and 10 light-green beads are found at every 20th position from beads to 1-181 and 201-381, respectively, and this pattern repeats. Data given below are for the state after 5x10$^4$ time units. (ii) Snapshots (with/without chromatin); red and green factors are found in
distinct clusters. (iii) Contact map (axes give bead numbers). The inset shows a high-resolution map of just binding beads in clusters (prepared as in Fig. 2Aiii). Here, red and green pixels mark contacts between two pink beads, or between two light-green beads, respectively. As each block
contains 10 binding beads (just less than the $\sim$12 typically found in a cluster in Fig. 1), as blocks alternate along the fiber, and as the two sets of bound factors assemble into distinct clusters, this fiber folds into a highly-organized structure -- which is reflected by the alternating colored squares along the diagonal. (iv) Rosettogram (pixels colored according to which high-affinity
beads are in the cluster). Again, this reflects the high level of organization (e.g., some ``perfect'' rosettes with 10 petals are present, and the $f_d$ is low).}

\newpage

\centerline{\includegraphics[width=10.cm]{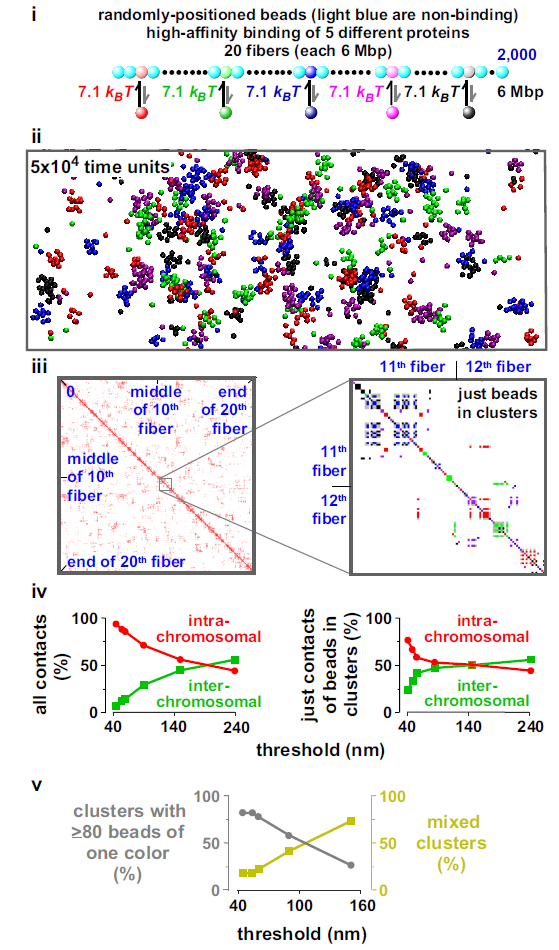}}
\captionof*{figure}
{{\bf Figure S5. Five different factors form distinct clusters when binding to cognate sites scattered randomly on 20 identical fibers.} 
MD simulations are those of Figure 2C. (i) Red, green, dark-blue, purple, and black factors (500 of each) bind (7.1 kBT) to five sets of cognate
sites scattered randomly along 20 identical fibers (each with 2,000 beads representing 6 Mbp). [Randomly scattering binding sites so that one in ~20 beads can bind a factor led in this case to 381, 385, 383, 437, and 416 binding beads in total for red, green, dark-blue, purple, and black
factors, respectively.] Data presented below were obtained after 5x10$^4$ time units. (ii) Snapshot (without chromatin for clarity); each factor tends to cluster with others of the same color (the center of
this image is presented in Fig. 2C). (iii) Contact map for all beads in every fiber (axes show positions; contacts made by every 100 adjacent beads on a fiber are binned). The inset shows a high-resolution map of just binding beads in clusters (as in Fig. 2Aiii) from bead 715 in fiber 11 to bead 1,123 in fiber 12; grey pixels mark contacts between beads of different colors, and
colored ones contacts between two beads of the indicated color. As 47\% non-white pixels are grey, most factors are present in clusters that contain only one color. (iv) The effect of the threshold used to define contacts (in nm) on the percentage of intra- and inter-chromosomal contacts between all beads (left), and between just binding beads (right). (v) The effect of the
threshold used to define contacts (in nm) on the percentage of clusters in which $\ge$80\% binding
beads are of one color, and in the other clusters (‘mixed’).}

\newpage

\centerline{\includegraphics[width=10.cm]{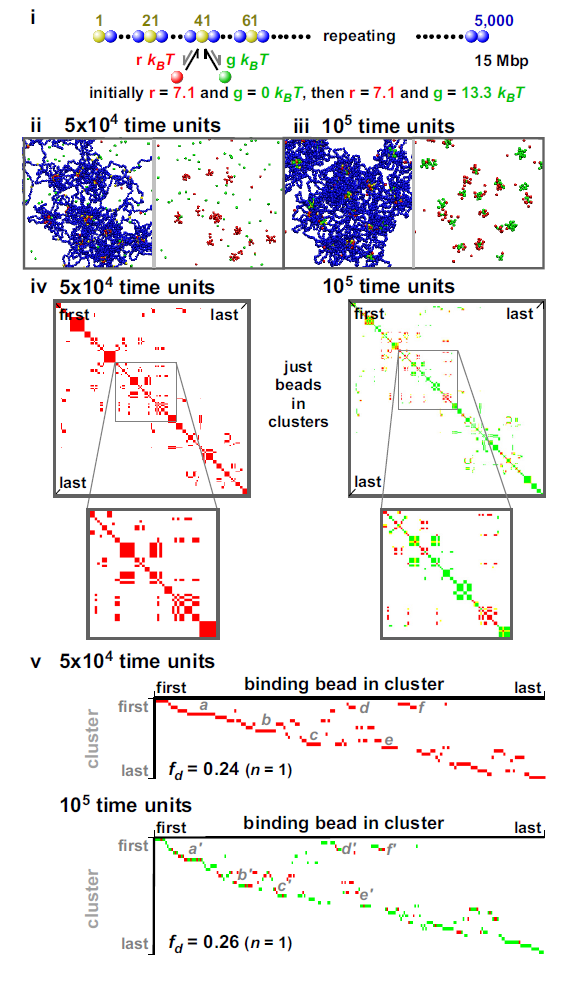}}
\captionof*{figure}
{{\bf Figure S6. Evolution of one type of cluster into another.} 
Conditions as in Figure 1, with exceptions indicated. (i) Overview. Red and green factors are present (250 of each); yellow beads are found at every 20th position in a 5,000-bead fiber. Initially, red factors interact (7.1 kBT) with yellow beads, but green factors do not interact with any beads. After 5$\times$10$^4$ time units, green factors acquire affinity (13.3 kBT) for yellow beads (perhaps because they become ``phosphorylated''), and the simulation continues for another 5x10$^4$ units. (ii,iii) Snapshots (with/without chromatin). The system evolves first into one containing only clusters of bound red factors, and -- once green factors start binding with higher affinity – red-green (and pure green ) clusters develop. (iv) Contact maps of just binding beads in clusters prepared as for the inset in Figure 2Aiii, where contacts are scored without binning if bead centers lie 90 nm apart, and binding beads are treated as if they possess the color of the nearest factor bound to the fiber. Using this
coloring scheme, red, green, and yellow pixels mark contacts between two red beads, between two green beads, and between a green and red bead, respectively. Insets show zooms of indicated
regions. After 5$\times$10$^4$ units, only red pixels are seen (as only red factors are bound in clusters, and the green factors are non-binding). After 105 units, green pixels predominate. Note that the
general patterns seen at the two times are similar; this is because once clusters of red factors appear, the general structure persists after the switch as red factors in a cluster are replaced by
green ones. (vi) Rosettograms (pixel corresponds to binding sites, their colors depict those of the nearest bound factor). After 5x10$^4$ time units, only red factors are in clusters; after 105 units, green pixels predominate – with red and green factors sometimes being found in one cluster
(giving pixels of different colors in one row). Many clusters also persist from one time to the next (reflected by the pattern of sets of contacts a-f being similar to that of sets a'-f').}

\newpage

\centerline{\includegraphics[width=10.cm]{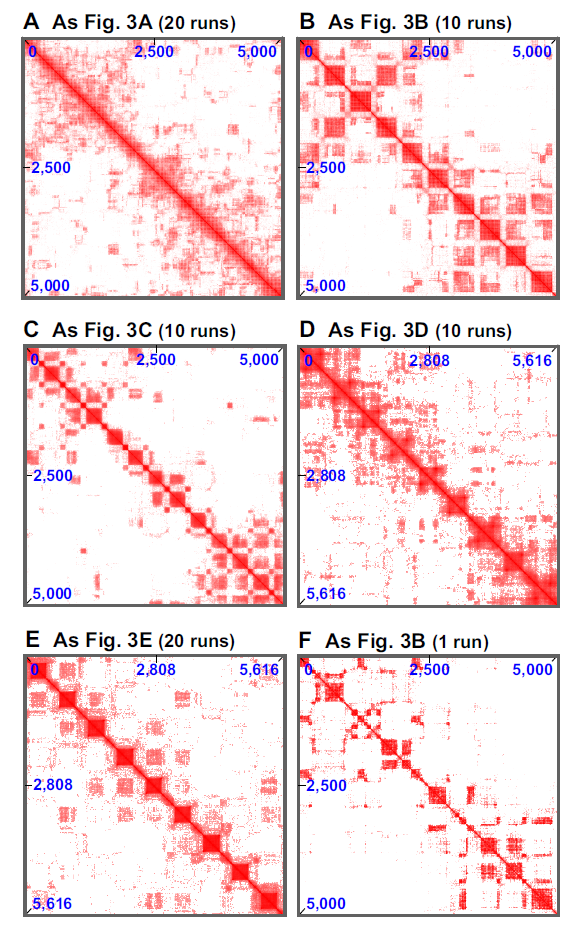}}
\captionof*{figure}
{{\bf Figure S7. Contact maps supporting Figure 3.}
(A-E) Truncated contact maps were presented in Figure 3; complete ones are given here. These
are averaged over the number of runs indicated.
(F) Complete contact map for one run using the conditions in Figure 3B; the off-diagonal blocks
(representing inter-domain interactions) visible here contribute only weakly to the population
average in (B).}

\newpage

\centerline{\includegraphics[width=10.cm]{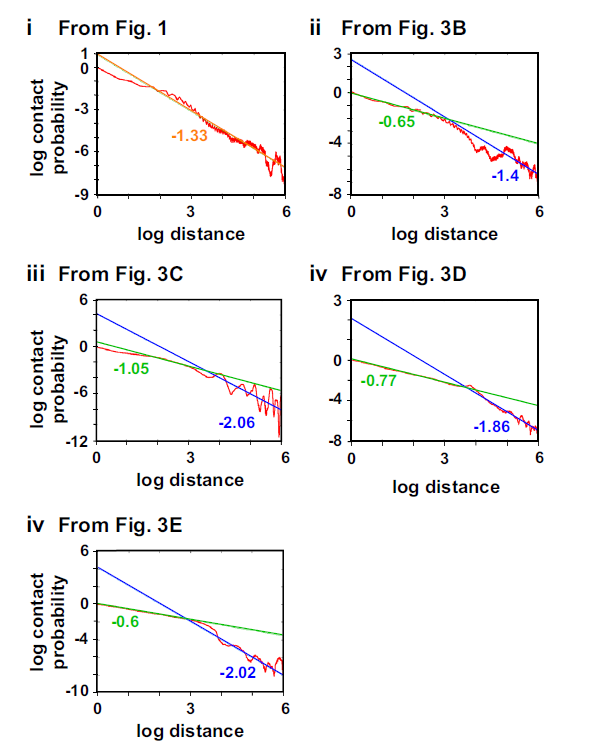}}
\captionof*{figure}
{{\bf Figure S8. Contact probability as a function of distance (bead number) along the different fibers illustrated in Figure 3 (red curves).} Straight lines indicate fits using the exponent indicated (brown line and exponent -- fit of entire curve; green line and exponent -- fit over short distances and so within a domain; blue lines and exponents -- fits over longer distances and so between domains). The effective exponent depends on distance (intra-domain versus inter-domain) and conditions; similar conclusions were reached by Barbieri et al. (2009).
}

\newpage

\centerline{\includegraphics[width=8.cm]{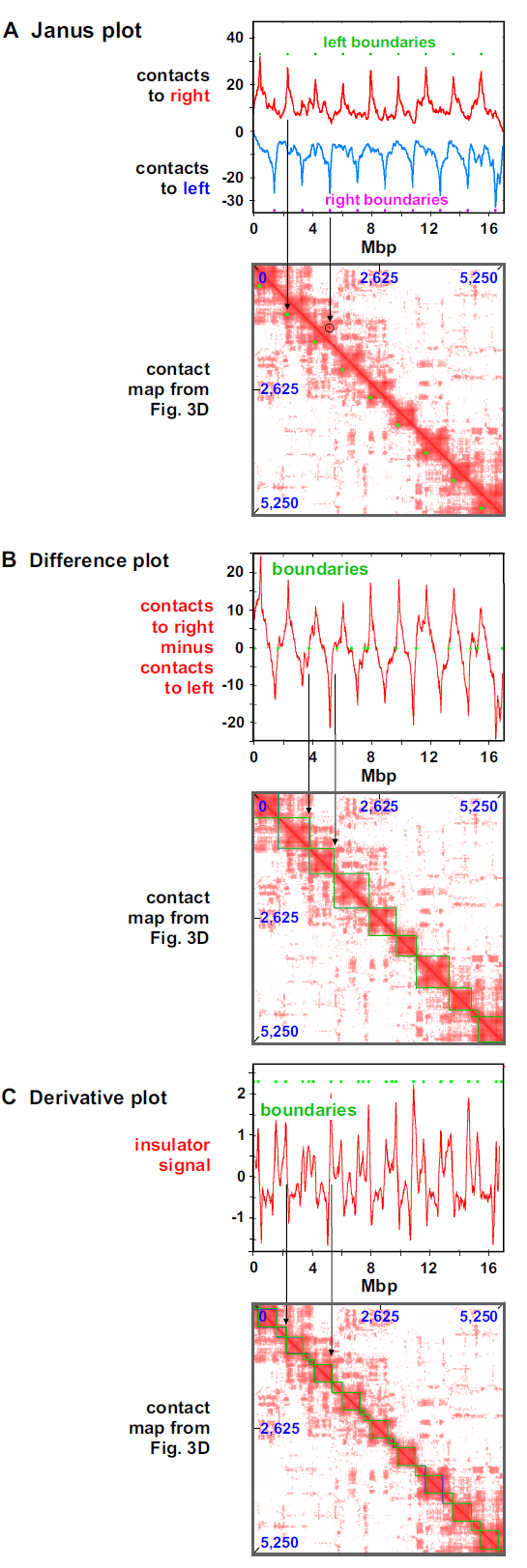}}
\captionof*{figure}
{{\bf Figure S9. Identifying loops and boundaries in contact maps.}
This Figure shows examples of how loops and boundaries can be identified using data in the contact map in Figure 3D (reproduced in each case below each of the 3 plots, with vertical arrows indicating related positions).
(A) ``Janus'' plot. This signal is proportional to the number of contacts made by each bead to the right (red curve) and to the left (blue curve). In the top plot (red), green circles identify peaks;
these mark beads at the left boundary of each loop. In the bottom plot (blue), purple circles identify the bottom of valleys; these mark beads at the right boundary of each loop. The coordinates of the left and right tethers found in this way can be used to identify contacts
corresponding to the base of the loop (these are shown as green dots or circles in the contact map).
(B) Difference plot. This shows the difference in contacts made by every bead (binned, with 7 beads/bin) to the right and to the left (i.e., the blue curve in A is subtracted from the red curve in A, see Supplementary Information; this plot is analogous to the one used by Dixon et al., 2012).
When the plot intersects zero with an upward derivative, it means that the pattern of contacts switches from contacts mainly to the left to mainly to the right (the behavior expected of a boundary).
(C) Derivative plot (see Supplementary Information, and Dixon et al., 2012) of data in (B). This can be viewed as a plot of an ``insulator'' signal, as now boundaries are identified with peaks (i.e., regions where the pattern of contact changes abruptly over a short genomic region).
}

\newpage
\centerline{\includegraphics[width=8.cm]{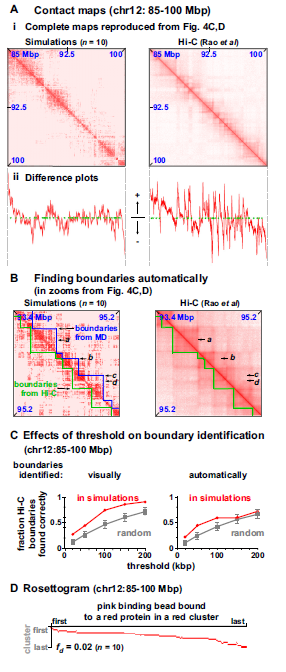}}
\captionof*{figure}
{{\bf Figure S10. Different ways of identifying boundaries in 15 Mbp on chromosome 12 (in GM12874) and the associated rosettogram.} 
Simulations were those used for Figure 4. (A) Contact maps and difference plots. (i) Contact maps are reproduced from Figures 4C,D. (ii)
Plot of the difference between contacts to the right and left for a given location on the simulated fiber (left; 7 kbp/bin) or the chromosome (right; 10 kbp/bin). Boundaries found by visually
inspecting the contact maps are shown by green dots; many (especially those close to the boundaries of the region analysed) – but not all – lie at or close to points where the plot intersects
zero with an upward derivative (plots prepared as in Fig. S9B).
(B) Finding boundaries by locating zeros in the difference plot with upward derivatives (see Fig. S9B and Supplementary Information) in the contact maps in Figure 4. Blue and green lines in the
zooms (same regions as in Figs. 4C, D) illustrate boundaries found automatically in the simulation and Hi-C data, respectively. Visual inspection indicates the algorithm is only partially successful at identifying boundaries. a: an obvious boundary in the simulation data that is missed
by the algorithm (this boundary is also seen in the Hi-C map, but is also missed by the algorithm). b: boundaries detected in both maps, but in the Hi-C map the algorithm places two boundaries very close to each other. c: the algorithm splits an obvious domain (which is seen in
the data from simulations). d: another boundary missed by the algorithm in both maps.
(C) Effects of threshold on correct prediction of boundaries (determined either by the difference plot aided by visual inspection, or automatically). A boundary is ``correctly'' predicted if it lies within a distance less than the threshold away from a boundary seen in the Hi-C data. For instance, 27 out of 36 boundaries (a fraction of 0.75) are correctly predicted by the difference
plot aided by visual inspection using a 100-kbp threshold. The grey line shows a control plot which gives the fraction of “correctly-predicted” boundaries found by scattering the same number of boundaries found in a simulation randomly throughout the genomic region analysed.
This procedure was repeated 100 times, and error bars in the random control denote the standard deviation. The difference between points on the two curves at most thresholds are highly significant (Table S2).
(D). Rosettogram for high-affinity beads in all 15 Mbp. Only pink binding beads that both bind red proteins and are in red clusters are considered. As grey and light-green binding beads are not considered, and as these are often found in long runs, the effects of such long runs on the
appearance of rosettograms are minimized; choice of only pink beads that bind red proteins further minimizes the effects of runs of adjacent pink beads. Many red pixels abut in one row, indicating the formation of many rosettes involving nearest-neighbor pink beds. The fd is also low, indicative of many local contacts and an ordered structure.}

\newpage

\centerline{\includegraphics[width=8.cm]{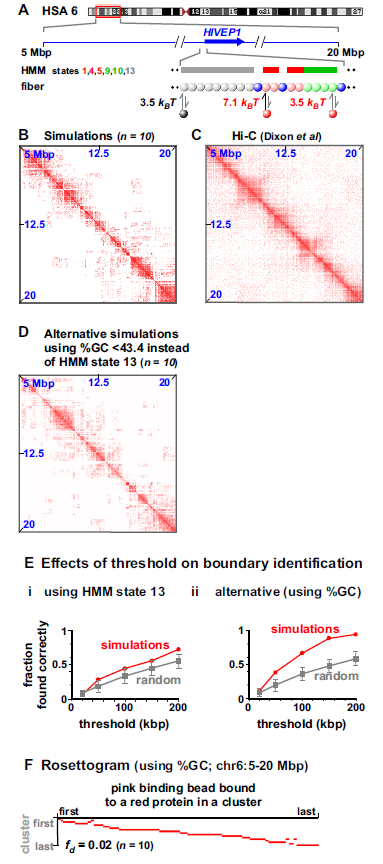}}
\captionof*{figure}
{{\bf Figure S11. Simulating 15 Mbp of chromosome 6 in H1-hESC cells.}
(A) Overview. The ideogram (red box gives region analyzed) and HMM track (colored regions reflect chromatin states) are from the UCSC browser; the zoom illustrates the HIVEP1 promoter.
Beads (1 kbp) are colored according to HMM state (blue -- non-binding, $n$= 3,890; pink -- states 1+4+5, $n$= 167; light-green – states 9+10, n = 723; grey -- state 13, $n$= 10,238). Red factors ($n$=300) bind to (active) pink and light-green beads with high and low affinities, respectively; black
(heterochromatin-binding) proteins ($n$= 3,000) bind to grey beads.
(B,C) Contact maps (7 and 20 kbp binning for simulation and Hi-C data, respectively). (D) Contact map obtained using alternative simulations, in which grey beads were selected using \%GC $>$43.4 (instead of HMM state 13). [This \%GC gives the same number of grey beads as the use of HMM state 13.] The overall pattern is similar to that seen in (B). (E) Effects of threshold on correct prediction of boundaries (determined by difference plots
aided by visual inspection). A boundary is ``correctly'' predicted if it lies within a distance less than the threshold away from a boundary seen in the Hi-C data. The grey line shows a control plot which gives the fraction of “correctly-predicted” boundaries ($\pm$ SD) found by scattering
randomly the same number of boundaries found in a simulation throughout the genomic region analysed. (i) Results obtained using the simulation illustrated in (A) and the contact map in (B).
(ii) Results obtained using the alternative set of simulations that give the contact map in (D), and a higher fraction of correctly-identified boundaries. The difference between points on the two curves at each threshold are now highly significant (see Table S2). 
(F). Rosettogram for all high-affinity beads in the 15 Mbp, prepared using the alternative data set that gave the contact map in (D). Only pink binding beads that both bind red proteins and are in clusters are considered. As grey and light-green binding beads are not considered, and as these are often found in long runs, the effects of such long runs on the appearance of the rosettogram are minimized; choice of only pink beads that bind red proteins further minimizes the effects of runs of adjacent pink beads. Many red pixels abut in one row, indicating the formation of many rosettes involving nearest-neighbor pink beds. The fd is also low, indicating many local contacts and an ordered structure.}

\newpage

\centerline{\includegraphics[width=8.cm]{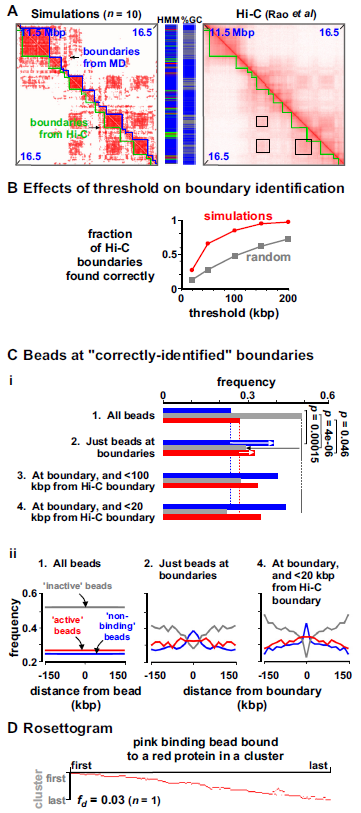}}
\captionof*{figure}
{{\bf Figure S12. Some properties of structures seen in simulations of chromosome 19 in GM12874 cells.} Simulations were those used for Figure 5.
(A) Zooms of contact maps reproduced from Figure 5C and D with added boundaries (21 and 20 kbp binning for data from simulations and Hi-C). Boundaries were determined using the ``difference'' plot aided by visual inspection (simulations -- blue lines; Hi-C -- green lines). Tracks
between zooms (HMM states and \%GC, colored as in Fig. 5A) show there is only partial correlation with domains in data from both simulations and Hi-C. Dashed rectangles in the Hi-C
map mark (off-diagonal) blocks of distant contacts seen in both sets of data.
(B) Effects of threshold on correct identification of boundaries (determined by visual inspection
of the whole chromosome). A boundary is ``correctly'' predicted if it lies within a distance less than the threshold away from a boundary seen in Hi-C data. The grey line shows a control plot which gives the fraction of ``correctly-determined'' boundaries found by scattering randomly the same number of boundaries found in a simulation throughout the genomic region analysed. Error
bars ($\pm$SD) in the random control are smaller than the square symbols and so cannot be seen, and the difference between points on the two curves are highly significant (typically $p <10^{-6}$, see Table S2).
(C) “Correctly-identified” boundaries in the whole chromosome are rich in active (pink and light-green) beads, and poor inactive (grey) beads. The frequencies of blue, grey, and pink+light-green beads (collectively depicted here by red bars and curves) in different sets of beads were
calculated. Set 1: all beads. Set 2: Beads lying within 100 kbp of a boundary (identified manually as in Figure 5). Sets 3 and 4: The sub-sets of set 2 that also lie within 100 and 20 kbp of a
boundary identified in Hi-C data. (i) Beads at boundaries are rich in active (pink+light-green) and blue beads, and depleted of inactive (grey) beads (arrows; p values assessed assuming Poisson distributions). (ii) The frequencies of different beads (in sets 1, 2 and 4) in the 150 kbp
on each side of either each bead in set 1, or of boundaries in sets 2 and 4. Boundaries are rich in blue (blue curves) and active beads (red curves), and poor in inactive ones (grey curves).
(D). Rosettogram for all high-affinity beads (pink) in the chromosome that both bind red proteins and are in clusters. Considering a sub-set of beads here has various advantages. First, as grey and
light-green binding beads are not considered, and as these are often found in long runs, the effects of such long runs on the appearance of the rosettogram are minimized. Second, choice of
only pink beads that bind red proteins further minimizes the effects of runs of adjacent pink beads. Third, these restrictions allow us to include all relevant beads in the whole chromosome in
one plot. Many red pixels abut in one row, indicating the formation of many rosettes involving neighboring pink beads. The $f_d$ is also very low, indicating an ordered structure.}

\end{document}